REVIEW ARTICLE

# 4D fabrication of shape-changing systems for tissue engineering: state of the art and perspectives

**Lorenzo Bonetti**[1] · **Giulia Scalet**[1]

Received: 23 April 2024 / Accepted: 30 July 2024

## Abstract
In recent years, four-dimensional (4D) fabrication has emerged as a powerful technology capable of revolutionizing the field of tissue engineering. This technology represents a shift in perspective from traditional tissue engineering approaches, which generally rely on static—or passive—structures (e.g., scaffolds, constructs) unable of adapting to changes in biological environments. In contrast, 4D fabrication offers the unprecedented possibility of fabricating complex designs with spatiotemporal control over structure and function in response to environment stimuli, thus mimicking biological processes. In this review, an overview of the state of the art of 4D fabrication technology for the obtainment of cellularized constructs is presented, with a focus on shape-changing soft materials. First, the approaches to obtain cellularized constructs are introduced, also describing conventional and non-conventional fabrication techniques with their relative advantages and limitations. Next, the main families of shape-changing soft materials, namely shape-memory polymers and shape-memory hydrogels are discussed and their use in 4D fabrication in the field of tissue engineering is described. Ultimately, current challenges and proposed solutions are outlined, and valuable insights into future research directions of 4D fabrication for tissue engineering are provided to disclose its full potential.

**Keywords** 4D fabrication · 4D printing · Biomedical engineering · Tissue engineering · Shape-memory polymers · Shape-memory hydrogels · Shape change

## Abbreviations

| | |
|---|---|
| AA-MA | Methacrylated alginate |
| BA | Butyl acrylate |
| BDE | Bisphenol A diglycidyl ether |
| BMSCs | Bone marrow stem cells |
| CVTE | Cardiovascular tissue engineering |
| DA | Decylamine |
| DIW | Direct ink writing |
| DLP | Digital light processing |
| DMPA | 2,2-Dimethoxy-2-phenyl acetophenone |
| FFF | Fused filament fabrication |
| GelMA | Methacrylated gelatin |
| GNPs | Graphene nanoplatelets |
| HA-MA | Methacrylated hyaluronic acid |
| HAp | Hydroxyapatite |
| hASCs | Human adipose stem cells |
| hECs | Human endothelial cells |
| hiPSC-CMs | Human-induced pluripotent stem cell-derived cardiomyocytes |
| hMSCs | Human mesenchymal stem cells |
| NIR | Near infrared |
| NSCs | Neural stem cells |
| OMA | Oxidized methacrylated alginate |
| PBE | Poly(propylene glycol) bis(2-aminopropyl ether) |
| PEG | Polyethylene glycol |
| PCL | Polycaprolactone |
| PCLDA | Polycaprolactone diacrylate |
| PEGDA | Poly(ethylene glycol) diacrylate |
| PGDA | Poly(glycerol dodecanoate) acrylate |
| PDMS | Polydimethylsiloxane |
| PHB | Polyhydroxybutyrate |
| PLA | Polylactic acid |
| PLLA | Poly(L-lactide) |
| PNIPAM | Poly(N-isopropyl acrylamide) |
| PU | Polyurethane |
| PVA | Polyvinyl alcohol |
| $R_f$ | Strain fixity rate |
| $R_r$ | Strain recovery rate |

✉ Lorenzo Bonetti
lorenzo.bonetti@unipv.it

[1] Department of Civil Engineering and Architecture (DICAr), University of Pavia, Via Ferrata 3, 27100 Pavia, Italy

| | |
|---|---|
| Sil-MA | Methacrylated silk fibroin |
| SLA | Stereolithography |
| SLS | Selective laser sintering |
| SME | Shape-memory effect |
| SMC | Shape-memory composite |
| SMH | Shape-memory hydrogel |
| SMP | Shape-memory polymer |
| SMPU | Shape-memory polyurethane |
| SOEA | Soybean oil epoxidized acrylate |
| tBA | *tert*-butyl acrylate |
| $T_g$ | Glass transition temperature |
| $T_m$ | Melting temperature |
| $T_{trans}$ | Transition temperature |
| TE | Tissue engineering |
| TEGDMA | Triethylene glycol dimethacrylate |
| TPU | Thermoplastic polyurethane |

## 1 Introduction

Since its introduction, almost four decades ago, tissue engineering (TE) has emerged as an alternative approach to tissue and organ transplantation, mitigating the critical shortage of these latter through the *in vitro* fabrication of functional biological structures [1]. Nowadays, it is possible to precisely control cells and the environment in which they are located, designing increasingly complex engineered tissue and organs capable of responding to specific clinical needs [2].

In this framework, the advent of 3D printing technology breathed new life to the TE field, enabling the production of complex, personalized structures and even living tissue constructs with exceptional precision and accuracy, while reducing material wastage and processing times [3–7]. Although the substantial progress made, an increasing demand for dynamic structures capable of recapitulating the complexity of living systems still exists.

4D fabrication holds significant promise in addressing this unmet need, opening new possible routes for fabricating complex, dynamic structures capable of responding and adapting to external stimuli in a programmed way. As a matter of fact, this field of research has continuously and exponentially grown over the past decade, as demonstrated by the rising number of annual publications on 4D fabrication (Fig. 1), amounting to 0 in 2010 versus more than 500 in 2023 (Scopus database). The concept of 4D fabrication dates back to 2013 [8], when the Tibbits' group at the Massachusetts Institute of Technology first talked about 4D printing as a process for the obtainment of structures with the ability to change shape, property, or functionality over time (i.e., the 4$^{th}$ dimension). Thanks to this revolutionary idea, the fabricated structures are no longer static objects, but are instead active objects whose transformation can be precisely engineered to respond to specific needs.

The implications of 4D fabrication in the field of TE (and more in general of biomedical engineering) are rousing, encompassing the ability of the 4D structures to dynamically change, self-transforming, self-adapting, and self-maturing after fabrication [9, 10]. This advancement holds the potential to enhance therapeutic outcomes and facilitate patient-specific treatments [11]. In this regard, 4D fabrication has the potential to revolutionize several fields of healthcare, including TE, drug delivery, medical devices/implants, and diagnostics [12]. For instance, in TE applications, 4D fabrication offers the possibility to fabricate constructs capable of transforming into the desired shape post-implantation, enhancing the integration into the host tissue, fostering cell proliferation and differentiation [11]. Likewise, in the field of drug delivery, 4D fabrication allows the obtainment of systems capable to administer the active principle in response to physiological or pathological cues, like pH or temperature variations, ultimately leading to targeted and more controlled delivery [13, 14]. In the field of diagnostics, 4D fabricated sensors and devices possess the ability to adapt to diverse biological conditions, thereby enhancing the precision and dependability in detecting biomarkers, pathogens, or other analytes [15].

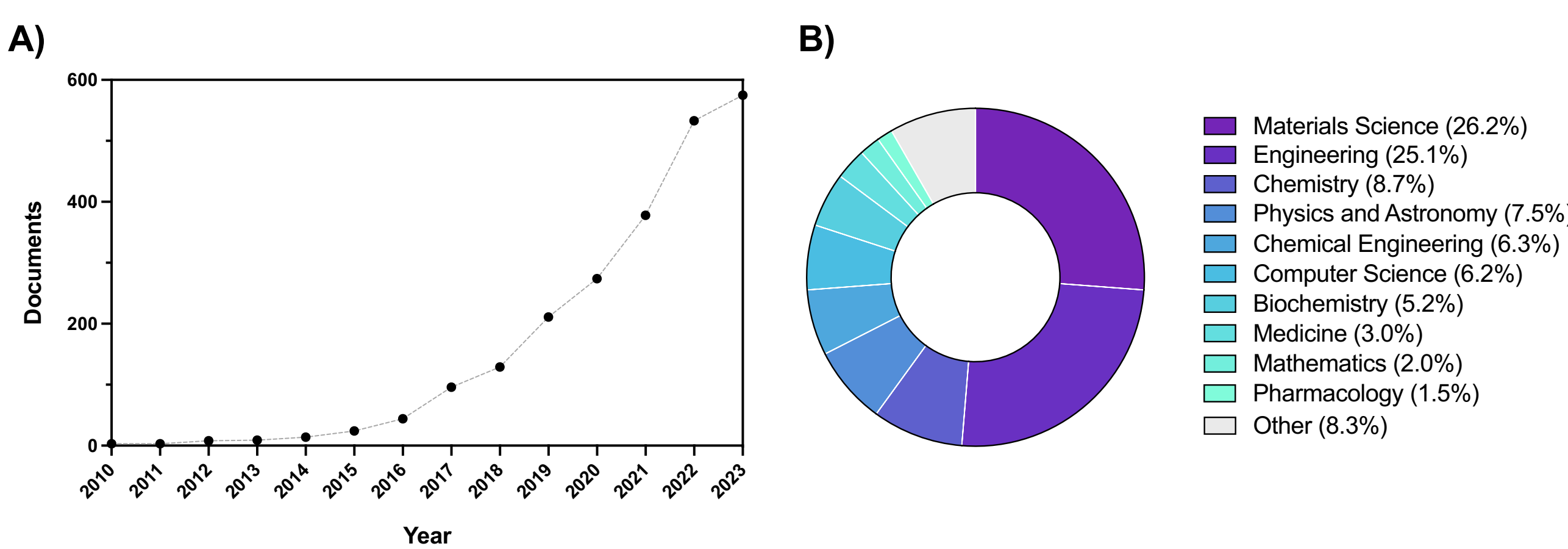

**Fig. 1** **A** Publications by year and **B** publications by subject area. Keywords: 4D fabrication, 4D biofabrication, 4D printing, 4D bioprinting. Source: Scopus database (2024/07/17)

4D fabrication mainly relies on three building blocks [16]:

i. the fabrication technology: the selected technology depends on the specifications of the structure to be fabricated (e.g., dimension, resolution), the material type (physical state, properties), and the material processing conditions (e.g., temperature, solvents).
ii. the stimulus-responsive material: stimuli-responsive materials can be classified into several sub-categories and the selection depends on their responsive ability, among others shape-memory, self-adaptability, or self-repair. The reader is referred to [17–22] for further reading.
iii. the stimulus: the stimulus is the trigger required to activate the response of the 4D fabricated structure. A plethora of different stimuli have been reported in the literature, ranging from temperature to light, solvents, and even a combination of different stimuli. The selection of the stimulus must be driven by the specific application, but is also related to the selected stimulus-responsive material.

In addition to the above-mentioned building blocks, mathematical modeling can further assist 4D fabrication [16].

This review explores the progress in 4D fabrication for advanced TE solutions. Typically, literature reviews in the field provide a general overview of 4D fabrication encompassing a wide range of materials and applications. However, they often lack a thorough examination of solutions obtained from shape-changing soft materials (i.e., soft materials responding to external stimuli by a shape variation). To bridge this gap, this review specifically focuses on 4D fabricated cellularized constructs obtained from these responsive soft materials, delving into three critical building blocks essential for this focus. First, the approaches to obtain cellularized constructs are introduced along with an overview of the main strategies for 4D fabrication of soft shape-changing systems, describing the materials that can be processed, the advantages and limitations, and the resolution of each technique, with the purpose to guide the reader throughout the choice of the best fabrication technique for the envisaged application. Then, the two main families of shape-changing soft materials, namely shape-memory polymers (SMPs) and shape-memory hydrogels (SMHs) are introduced, elucidating the mechanisms and ways of their shape transformation. Finally, an overview of the main applications of 4D fabricated cellularized constructs for TE purposes is introduced and the future perspectives in the field are highlighted. By concentrating on these key areas, our review provides a comprehensive and detailed perspective on the potential of 4D fabrication using shape-changing soft materials. We highlight the innovative solutions these materials offer for creating next-generation tissue engineering constructs, capable of dynamic and functional integration within biological systems.

## 2 4D fabrication approaches

In the last decades, the demand for increasingly complex scaffolds, medical devices, and products for TE purposes has driven the need for progressively more advanced fabrication technologies. Nowadays, both conventional and non-conventional fabrication approaches are used for the design of scaffolds and constructs [23]. Moreover, such approaches are often combined with additional micro- and nanofabrication methods to impart topographical cues and control cell-substrate interactions and cell fate [24].

For a better understanding of this work, we will provide an overview of the approaches to obtain 4D cellularized structures (Fig. 2). Moreover, an overview on the main techniques for the 4D fabrication of scaffolds/constructs (Table 1), also including several useful references for further reading, will be provided to support the reader's understanding of the subsequent works reviewed.

In accordance with Ionov [25], we will classify the approaches to achieve three-dimensional cellularized structures as follows (Fig. 2): (A) fabrication of non-vital structures, their shape change, and cell seeding; (B) fabrication of non-vital structures, cell seeding, and their shape change; (C) (bio)fabrication of vital constructs and their shape change. Hereafter, we will refer to the three approaches as 4D-A, 4D-B, and 4D-C, respectively. In Sect. 3.4, we will discuss how different fabrication techniques have been explored in the field of TE following 4D-A, 4D-B, and 4D-C approaches.

### 2.1 Conventional fabrication techniques

Conventional fabrication techniques can be distinguished into formative (i.e., molds) and subtractive (i.e., machining). Such technologies are generally easy to implement and do not require costly infrastructures, but they may require multiple steps (e.g., post processing), thus becoming time-consuming. Most of these techniques are limited to 2D or simple 3D structures and cannot control the geometrical features precisely [23].

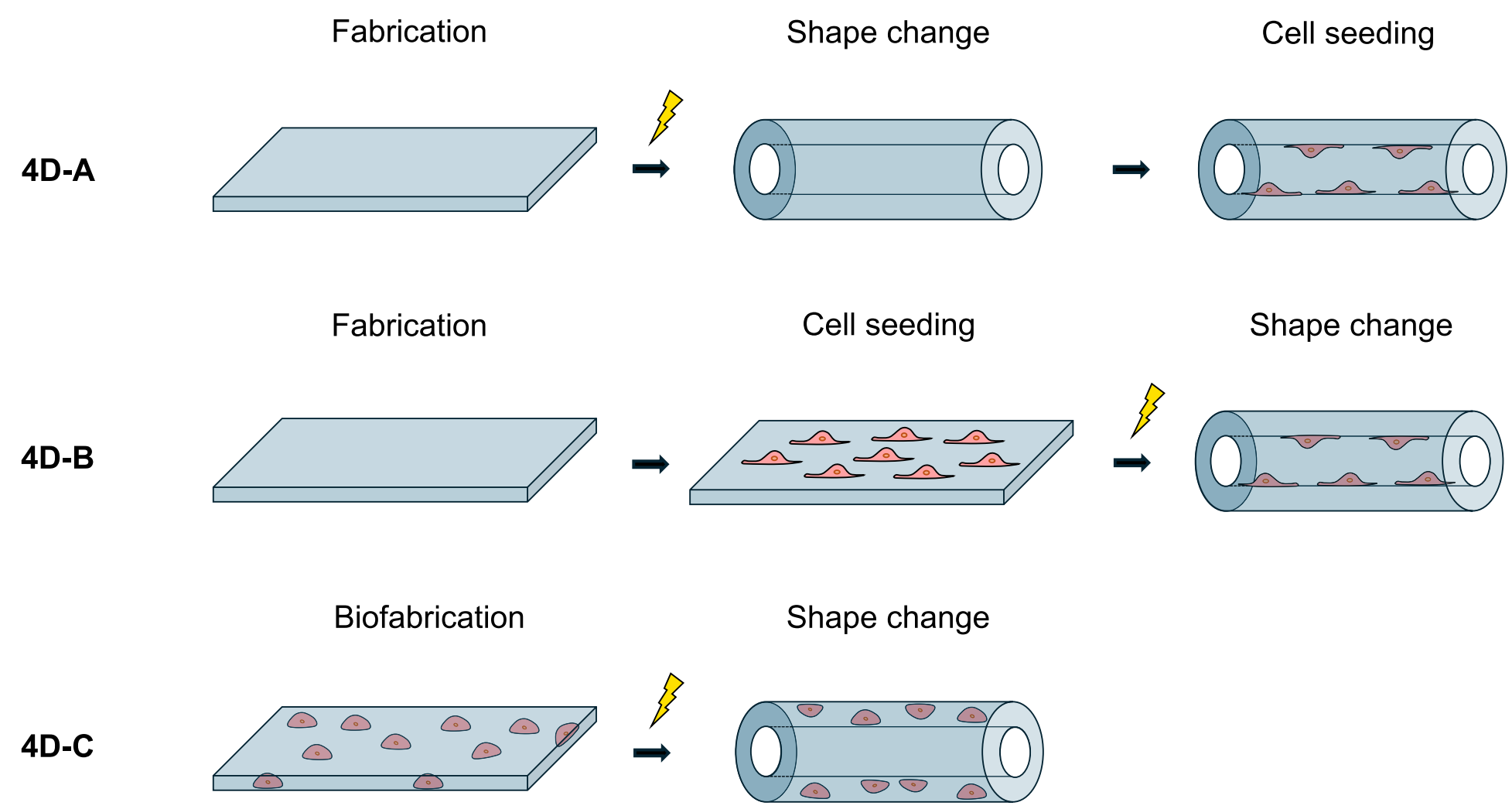

**Fig. 2** Scheme of 4D fabrication of cellularized constructs exploiting shape-changing materials: **4D-A** fabrication of a non-cellularized scaffold, shape change, and seeding with cells; **4D-B** fabrication of a non-cellularized scaffold, cell seeding, and shape change of the construct; **4D-C** biofabrication of a cellularized construct and shape change. Re-adapted from [25]

#### 2.1.1 2D structures

Solvent casting is a fabrication method based on the dissolution of one or more polymers (with or without plasticizer(s) addition) in a volatile solvent (e.g., ethanol, acetone, water) and subsequent pouring of the solution on a substrate. The subsequent evaporation phase leads to the solvent removal, resulting in the obtainment of a film [26, 27].

#### 2.1.2 3D structures

Injection molding is one of the most employed techniques for the production of objects starting from thermoplastic materials. This technique allows for the obtainment of three-dimensional polymer shapes and usually does not require additional finishing. The underlying concept of injection molding is straightforward. The thermoplastic polymer is heated until it transforms into a viscous melt. Subsequently, it is injected into a sealed mold, determining the desired shape of the object. Within the mold, the material undergoes cooling until it solidifies, then the mold is opened, allowing for the extraction of the final object [28].

#### 2.1.3 Porous scaffolds

Various methods including freeze-drying, solvent casting/particulate leaching, and gas foaming have been used to fabricate porous scaffolds.

Solvent casting/leaching is a simple process which involves leaching out solid particles from a polymer matrix. Specifically, particles (generally salt crystals) with a defined diameter are added to the polymer solution. Following solvent evaporation (e.g., via air-drying, vacuum-drying, or freeze-drying), the particles entrapped into the polymer matrix are leached out through immersion in a suitable solvent (e.g., water), generating a porous structure. Salt particles (e.g., sodium chloride) are mainly used, but sugar, sucrose, and starch, gelatin or paraffin microparticles have also been reported [29, 30].

Gas foaming is a widely employed method for the fabrication of porous scaffolds. The process is simple, and consists in the addition of a foaming agent (e.g., sodium bicarbonate) into an acidic polymer solution, generating an inert gas such as $N_2$ or $CO_2$. The porous structure is then achieved by removing the discontinuous phase (i.e., gas phase) from the continuous phase (i.e., polymer) [31, 32].

However, this technique may be plagued by issues in the control of the pore diameter, usually too large to favor cell proliferation. Thus, another foaming approach consists in introducing an inert gas ($CO_2$, $N_2$) into a melted, pressurized polymer, allowing to obtain scaffolds with better control of pore dimensions and homogeneity [29, 31, 32].

### 2.2 Non-conventional fabrication techniques

#### 2.2.1 Additive manufacturing

Despite conventional scaffold fabrication techniques have evolved in the last years, they are generally not useful when intricate and complex geometries are needed [33]. Moreover, conventional fabrication techniques are even plagued by the impossibility to answer the need of personalized or patient-specific geometries, often required in biomedical engineering applications.

Additive manufacturing has emerged as a powerful technology capable of taking up such challenges and contributing to advancement of several sectors. Particularly, it has revolutionized the biomedical field, leading to the production of personalized medical devices, implants, scaffolds, drug delivery platforms, actively contributing to advances in TE [34].

**2.2.1.1 Extrusion-based technology** Material extrusion relies on the use of a single material or a mixture of materials, commencing in either a liquid state or made into a viscous or amorphous consistency. The material is extruded through a nozzle tip to produce a continuous filament and deposited in a layer-by-layer fashion to generate the desired 3D structure [35].

Fused filament fabrication (FFF), often reported as fused deposition modeling (FDM), is undoubtedly the best-known and most investigated extrusion-based 3D printing technique. Such a technique has found extensive application in both industrial and laboratory environments given its simplicity: making use of a heated nozzle-based extruder equipped with a drive gear, a thermoplastic polymeric filament is melted and deposited in a layer-by-layer fashion, creating a 3D object (Fig. 3A) [35, 36]. Variations of the FFF process are often used, e.g., starting from pellets instead of filaments [37, 38].

Direct ink writing (DIW) refers to a 3D printing technique that makes use of pressure to extrude shear thinning fluids through a nozzle, by means of a computer-controlled print head, to layer-by-layer fabricate 3D structures (Fig. 3B). Three main types of DIW extruders have been reported in the literature, distinguishing into: i) pneumatic, ii) mechanical (i.e., piston and screw), and iii) solenoid-based extrusion systems. To date, DIW has been successfully investigated for printing of several classes of materials, e.g., metal particles, polymers, ceramics, and composites [36, 39].

**2.2.1.2 Light-assisted technology** Stereolithography (SLA) is a 3D printing technique used to generate three-dimensional objects in a layer-by-layer fashion by means of a photochemical process. Specifically, a laser is exploited to photo-crosslink liquid polymers and resins into solid or gel-like structures. In other words, photo-crosslinking occurs when the laser beam encounters the photo-curable material, causing it to bond together into a solid structure (Fig. 3C) [36, 40].

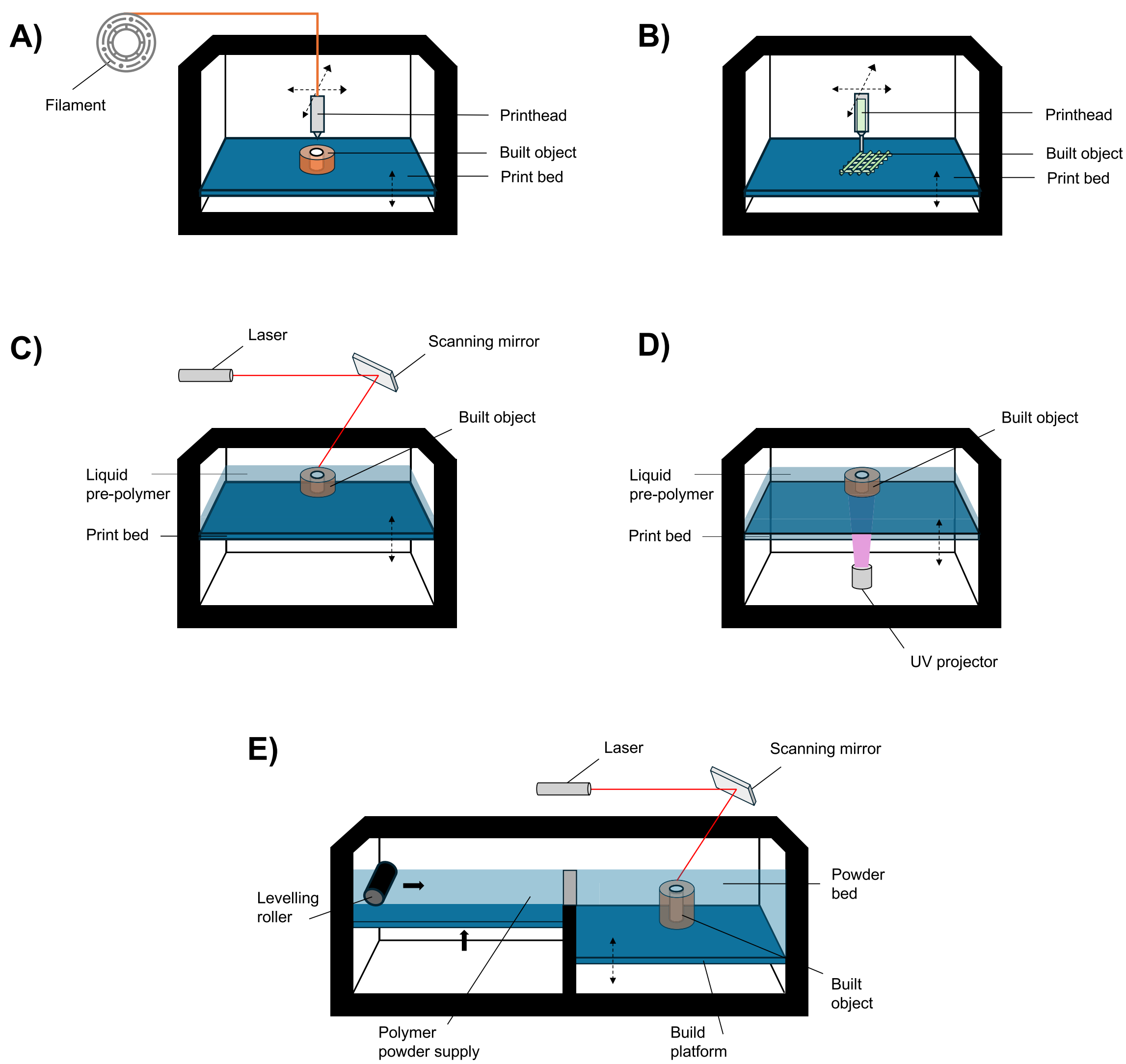

**Fig. 3** Additive manufacturing techniques employed for the 4D fabrication of cellularized constructs. Schematic illustration of **A** FFF, **B** DIW, **C** SLA, **D** DLP, and **E** SLS techniques

**Table 1** Fabrication techniques employed in the literature for the obtainment of cellularized structures

| Fabrication technique | Material(s) | Advantages/limitations | Resolution/porosity | Reference(s) |
|---|---|---|---|---|
| Solvent casting | Polymers in solution | Simplicity, low-cost, easy control of the film thickness, possibility to incorporate heat-sensitive molecules. / High amounts of solvent(s) and the long drying time | – | [42, 43] |
| Solvent casting/Particulate Leaching | Polymers in solution | Simplicity, low-cost. / Poor interconnection, irregular pore shape | Pore size: 100–500 $\mu$m (> 90 % porosity) | [29, 30] |
| Gas foaming | Polymers in solution/ melted polymers | Simplicity, low-cost, no need of organic solvents. / Poor control of pore size and interconnection shape | Pore size: 50–2000 $\mu$m (> 90% porosity) | [29, 31, 32] |
| Fused Filament Fabrication (FFF) | Thermoplastic polymers and composites | Medium-fast, simplicity of use, low cost. / Low resolution | 100 - 400 $\mu$m | [35, 36] |
| Direct Ink Writing (DIW) | Viscous fluids (e.g., polymers, hydrogels) | Medium-fast, materials versatility, good layers bonding. / Tuning of the rheological performance | 100 $\mu$m | [36, 39] |
| Stereolithography (SLA) | Photocurable materials (e.g., resins, hydrogels) | Medium-fast, High resolution and surface finish. / High costs, post processing (washing, curing) | 25–300 $\mu$m | [36] |
| Digital Light Processing (DLP) | Photocurable materials (e.g., resins, hydrogels) | Fast, high resolution, low cost. / Post processing (washing, curing) | 0.6–200 $\mu$m | [36] |
| Selective laser sintering (SLS) | Polymer powders | Fast, low cost, no support needed. / Grainy and porous finish | 1–150 $\mu$m | [36, 41] |
| Electrospinning | Polymers in solution or viscous state | Fast, easy control of fiber diameter, porosity, and pore size. / Possible fiber instability post-fabrication | 100 nm to 1 $\mu$m (80–95%) | [42, 43] |
| Photolithography | Photoresist | High resolution, fast. / Expensive equipment, need for cleanroom, only works on flat substrates | 100 nm (15 nm ultra-high resolution) | [48] |
| Replica molding | Photo and thermal curable polymers | Works on rigid and soft substrates, even on large and non-planar surfaces, low costs. / Wear or dissolution of the PDMS mold may occur due to use of solvents | 30 nm | [48] |
| Hot embossing | Thermoplastic polymers | High resolution, no solvent(s) needed, low cost. / Need of relatively high temperature and pressure | 5–10 nm | [45] |

SLA has found widespread use in the field of 4D printing, i.e., the combination of 3D printing and stimuli-responsive materials, towards the obtainment of structures whose shape, property, and functionality evolve with time (4$^{th}$ dimension). A large variety of stimuli-responsive materials, like SMPs and liquid crystal polymers have been investigated in the fabrication of structures capable of undergoing 4D shape changes [36].

Digital Light Processing (DLP) is another 3D printing technique that exploits light in the fabrication process.

Differently from SLA, DLP uses a digital light pattern (instead of a laser beam) to photo-crosslink liquid polymers and resins in a layer-by-layer process (Fig. 3D). Due to greater speed than SLA, DLP has established itself as a promising 3D printing technique for rapidly fabricating complex 3D (and 4D) structures, with micro- to nanoscale structural features [36].

Selective Laser Sintering (SLS) uses a laser beam to melt thermoplastic materials spread by a roller in the form of a tightly compacted powder onto a print bed. For each layer, heat generated by the laser selectively melts the powder under the control of a scanner system. Once the layer is built, the print bed is moved down (by the thickness of the newly fabricated layer) by a piston to accommodate a new layer of powder (Fig. 3E). The temperature inside the fabrication chamber is kept just below the melting temperature ($T_m$) of the thermoplastic polymeric powder, this way melting is achieved by a slight increase of temperature provided by the heat from laser. Powders with particle size ranging from tens to hundreds of microns are generally used for SLS [36, 41].

#### 2.2.2 Electrospinning

Electrospinning is a fabrication technique which allows for the obtainment of polymeric fibers through electrostatic forces applied on electrically-charged polymer(s) in solution or in a viscous state. The electrospinning set-up is mainly made up of four parts: i) a syringe (loaded on a syringe pump) containing the polymer solution, ii) a power supply, iii) a metallic needle or spinneret, and iv) a metallic collector (with different possible morphologies) [42, 43]. Scaffold fabrication is possible by connecting the spinneret and the collector to the power supply, leading to a potential difference among the two electrical terminals. Such a potential difference allows the polymer solution to flow from the spinneret to the collector, generating fibers with a diameter in the nano to micro scale.

### 2.3 Micro- and nanostructured substrates

Remarkable advances in micro- and nanoscale surface patterning technologies have opened up new possibilities and studies focusing on cell-surface interactions. In this regard, several methods allow the obtainment of surface topographies featuring geometrically-controlled micro- and nanopatterns (e.g., channels, pillars, and pits) capable of acting on cell-substrate interactions and guiding cell fate (e.g., adhesion, proliferation, and differentiation) [24, 44].

Photolithography is an optical means of transferring a pattern on a substrate. Typically, a silicon wafer serves as the substrate. A photoresist is then poured on the substrate and patterns are generated by exposing the substrate to high-intensity UV irradiation through a patterned photo mask, a film permitting UV light transmission only through the unmasked regions. As the last step, etching and dissolution in an appropriate solvent (developer solution) lead to removal of selected areas of the film [24].

Soft lithography is a family of techniques, that can be divided into: i) microcontact printing, ii) micromolding in capillaries, iii) microtransfer molding, iv) replica molding, and v) solvent-assisted micromolding [45]. For the purpose of this review work, only replica molding will be described. In fact, it is the most investigated patterning technology on SMPs in TE, within the family of soft lithographic techniques. For further details on other soft lithographic techniques, the reader is referred to [45]. The process of replica molding starts with the fabrication of a micro- or nanopatterned substrate on a silicon wafer via photolithography. This pattern then serves as a template for obtaining a mold, generally using polydimethylsiloxane (PDMS). Subsequently, a pre-polymer is cast onto the PDMS mold and micro/nano patterns are obtained on the polymer substrate via photo or thermal curing [24, 44, 45].

Hot embossing is a widely employed micropatterning technique based on the use of thermoplastic polymers. The working principle is quite straightforward: the work material (thermoplastic polymer) is placed on the lower plate of a press and the master (or mold) is attached to the upper plate of the press. The polymer is then heated above its glass transition temperature ($T_g$) and the master is pressed onto it. After cooling, the polymer replica is detached from the master [24, 46, 47].

## 3 Shape-changing materials

Different approaches have been described in the literature for the obtainment of a shape change. In this regard, it can be useful to classify these approaches according to the actuating system, thus distinguishing into i) shape-memory polymers (SMPs), ii) shape-memory hydrogels (SMHs), and iii) others. In the following sections, these systems will be overviewed focusing, for each of them, on the activating stimulus (e.g., temperature, ions).

### 3.1 Shape-memory polymers (SMPs)

SMPs are a class of intelligent polymers, capable to undergo defined—or programmable—shape changes (e.g., bending, stretching/contraction, or twisting) when exposed to an external stimulus [49]. Based on the nature of the external stimulus, SMPs have been classified as thermo-, light-, solvent-, and redox-sensitive [50].

For the purpose of this review, thermo-responsive SMPs, i.e., those activated via direct or indirect (e.g., light- [51] or magnetic-assisted [52]) heating, will be mainly discussed, both because they are the most studied in the literature panorama and for the ease in applying the thermal stimulus. This is even more striking in the biomedical field, where body temperature can be exploited as a trigger to activate the shape change. For further details on other activating stimuli, the reader is referred to [21, 22, 50, 53, 54]. In this framework, it is possible to distinguish between (i) one-way, (ii) multiple-way, and (iii) two-way SMPs (Fig. 4), depending on the type of shape-memory effect (SME). One-way SMPs can return from a temporary shape (i.e., a deformed state obtained through a "programming" process) to their permanent shape (i.e., the original shape obtained after their processing) upon heating (Fig. 4A). Multiple-way SMPs can recover their permanent shape from two or more temporary shapes upon heating (Fig. 4B). However, both one-way and multiple-way SMPs are not capable to return to their temporary shape after heating. Conversely, two-way SMPs display the capability of reversible, bidirectional movement between two different configurations—or programmable states—on the application of heating/cooling stimuli (Fig. 4C) [49, 54].

The temperature at which the shape recovery occurs is usually referred to as thermal transition temperature ($T_{trans}$), which can be either a glass transition temperature ($T_g$) or a melting temperature ($T_m$) depending on the type of SMP. The molecular mechanism underlying the SME, not covered by this review, is a defined polymer network architecture consisting of netpoints (i.e., covalent bonds or intermolecular interactions) and switching domains (i.e., reversible covalent bonds or crystallization/vitrification domains). For a comprehensive review of these aspects, the reader is referred to [21, 49, 55].

SMPs have attracted increasing attention over the last decades, due to their low costs, ease of processability, and flexibility typical of polymeric materials [56]. These properties, coupled with the possibility to arouse programmable and fine-tuned shape changes, have opened the floodgates to the application of SMPs in different research fields, and in particular for a variety of biomedical applications including medical devices (e.g., vascular stents, craniofacial plates), drug delivery vehicles, and scaffolds/constructs for tissue engineering and regenerative medicine [57]. In this regard, in addition to the need of the SMPs to exert their responsive behavior, further requirements -i.e., being non-toxic, cell adhesive, and biocompatibile- are needed to avoid inflammatory responses and support cell functions [21, 50, 53, 58]. Among the fabrication techniques described before in this review (see Sect. 2.2), FFF and DLP are probably the most employed ones for the fabrication of structures from SMPs and, more in general, from stimuli-responsive hydrogels.

### 3.2 Shape-memory hydrogels (SMHs)

Hydrogels are a noteworthy class of materials constituted by water-insoluble, 3D networks of polymer chains capable of retaining large amounts of water [59]. SMHs represent a subclass of the hydrogels' family capable of undergoing

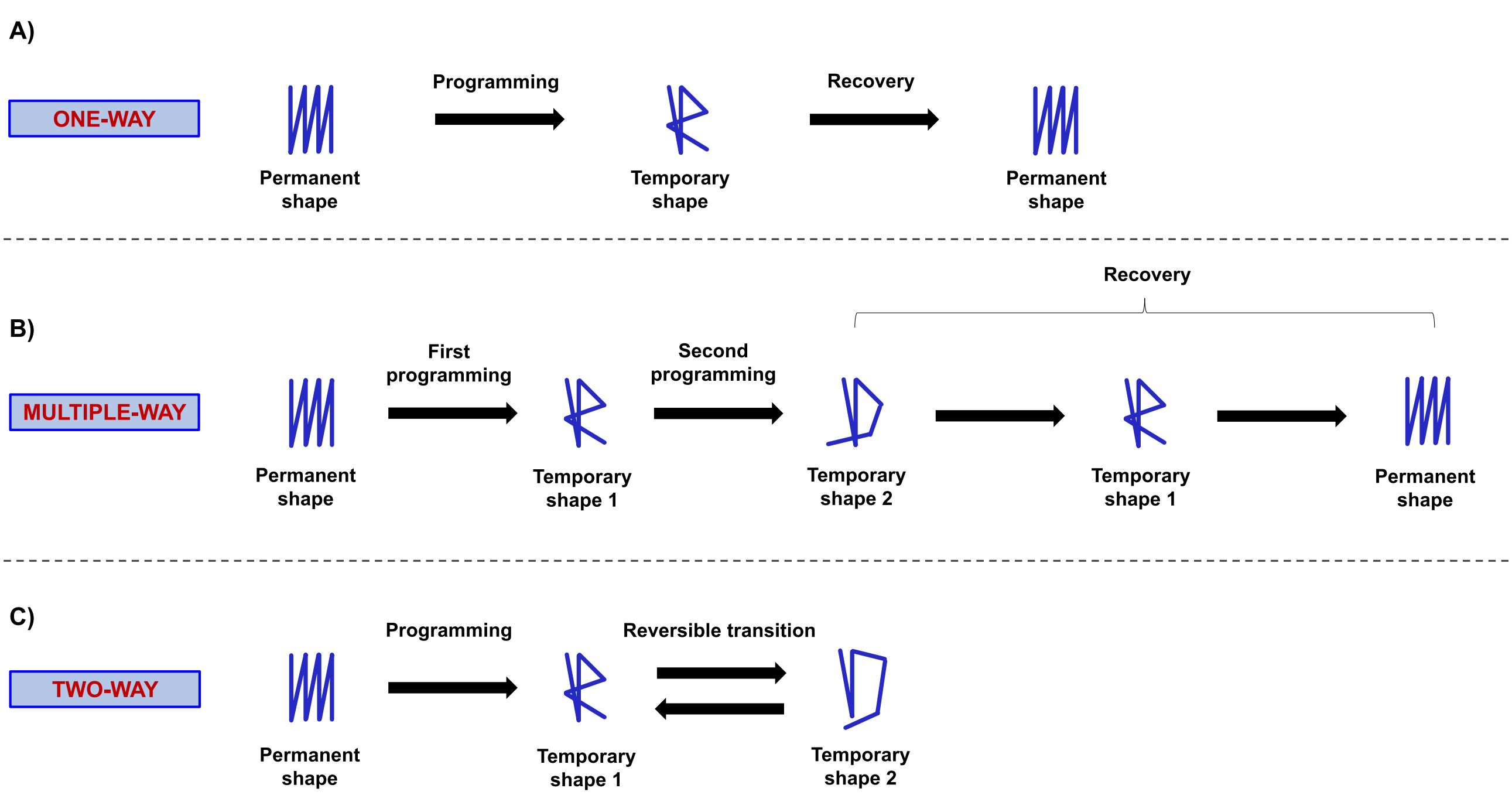

**Fig. 4** SME of **A** one-way, **B** multiple-way, and **C** two-way SMPs and SMHs. Note that, to clarify the concept of multiple-way SME, the second row of this figure represents a triple SMP/SMH having one permanent and two temporary shapes. Re-adapted from [54]

defined shape changes upon exposure to external stimuli. Their actuation mechanism is mainly based on their ability to selectively swell and deswell upon exposure to a plethora of different stimuli, including solvent type, pH, temperature, and light [19, 60–62]. As for SMPs, it is possible to distinguish between one-way, multiple-way, and two-way SMHs according to the SME (Fig. 4). Complex shape changes (e.g., folding, twisting) are commonly achieved introducing anisotropies in the 3D structures. In this regard, anisotropic structures are generally multilayered structures, where the different swelling rates in the layers can be achieved using different materials (i.e., having different swelling degrees) or different crosslinking densities [36, 60, 62].

Conversely to the majority of solid-state SMPs, in which cells can only be seeded on the surface of the materials (i.e., approached 4D-A, 4D-B), in SMHs cells can be uniformly dispersed within the same gels [39]. For this peculiar ability, SMHs are promising candidates for 4D biofabrication (i.e., approach 4D-C). Given their favorable properties, among which biocompatibility, biodegradability, and biomimetic nature, SMHs may be incredibly attractive for a wide variety of biomedical applications, ranging from TE purposes to drug/cell delivery and 3D/4D (bio)fabrication [36, 59].

### 3.3 Other approaches

Other approaches have been reported in the literature to obtain a shape change. Such approaches, hereafter only mentioned, will not be covered in this work. For further information, the reader is referred to [25].

A first possible approach is based on exploiting cell contraction forces. In fact, cells adhered on a substrate may be able to exert forces on the substrate itself, leading to self-folding constructs. Following this concept, Kuribayashi-Shigetomi and co-workers [63] reported a method to generate self-folding 3D constructs exploiting the principle of origami folding and cell traction forces. In details, two or more micro-patterned microplates, connected by flexible joints, were fabricated from parylene (poly(p-xylene) polymer) and coated with fibronectin. The cells seeded on such microplates exerted the traction forces needed to fold the 2D microstructures into complex shapes and structures (Fig. 5A). This technique laid the foundation for the next generation cell-based biohybrid medical devices (e.g., grafts or constructs [64]), and for advances in the fields of cell biology under flexible and configurable 3D environments.

Another interesting approach is based on spontaneous deformation based on internal stresses generated in the materials. Specifically, multi-layered thin films having different mechanical properties have been reported to spontaneously self-fold, transforming from 2D to 3D geometries. On this topic, Teshima and co-workers [65] fabricated micro-patterned films based on silk fibroin hydrogel and poly(cholro-p-xylene) (parylene-C) capable to autonomously self-fold into cylindrical shapes according to the different strain gradients present in the films (Fig. 5B). Interestingly, the folding extent was revealed to be dependent on the film thickness, and different 3D cell-laden constructs were fabricated following this approach. These results can open the gates to the fabrication of 3D bio-interfaces with countless biomedical outcomes, ranging from the reconstruction of functional tissues to implantable tissue grafts.

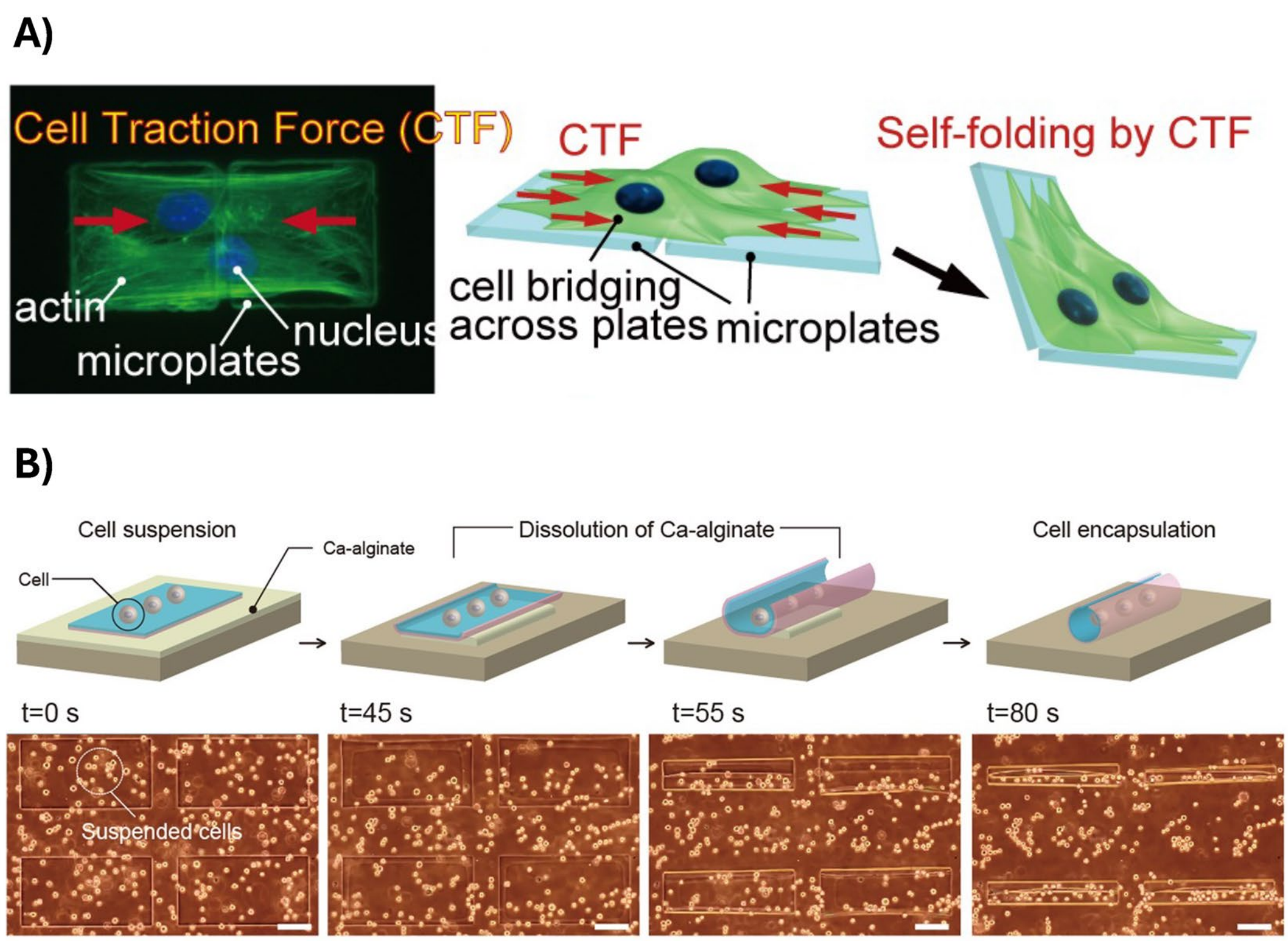

**Fig. 5** Approaches for the obtainment of shape changes not based on SMPs and SMHs. **A** Cell contraction forces: cells adhere and stretch across two micro-fabricated parylene microplates and the exerted cell traction forces generate the folding from 2D to 3D microstructures. Green = actin, blue = nuclei. Reproduced from [63]. **B** Internal stresses: time-lapse representation and optical images displaying cell encapsulation inside self-rolling structures according to their strain gradients. Reproduced from [65]

Similar approaches have also been reported, for instance exploiting photo-crosslinked polyethylene glycol bilayers [66] or polysuccinimide/polycaprolactone bilayers [67].

The potentialities of approaches just described are countless, and rely on the possibility to achieve a shape change directly after the fabrication process (i.e., 4D-B approach). Therefore, cells seeded on the resulting structures can grow and adapt to the dynamic environment that occurs during the shape transformation process.

### 3.4 4D fabrication in the biomedical field

Hereafter, an overview of the progress made in the field of 4D fabrication will be provided. Specifically, the main studies dealing with biomedical applications of 4D fabrication will be reported (Tables 2, 3), distinguishing them according to the specific application (i.e., scaffolds and cell culture surfaces) or body district of application for TE purposes. In each section, SMPs and SMHs will be discussed individually.

#### 3.4.1 Scaffolds and cell culture surfaces

**3.4.1.1 SMPs** The first works combining cells and SMPs towards 4D fabrication mainly dealt with the fabrication of scaffolds and cell culture surfaces, unveiling the role of the shape change on the cellular functions (e.g., viability, cytoskeletal/nuclear re-arrangement, differentiation, internalization). Several polymers and cell types have been tested (Table 2), also elucidating the effect of the sole (i.e., not coupled with a SMP) temperature change on the cell functions.

Polycaprolactone (PCL) and relative co-polymers are probably the most employed SMPs. Le and co-workers [68] were among the first to report UV-crosslinked PCL to fabricate surfaces useful to guide the alignment of human mesenchymal stem cells (hMSCs). UV-crosslinked PCL displayed a $T_m$ close to body temperature and excellent strain fixity ($R_f$) and strain recovery ($R_r$) rates (99 and 98 %, respectively). Moreover, the change in surface topography from microarray (3 x 5 $\mu$m array) to flat, induced by temperature increase (from 28 to 37 °C), led to a change in cell alignment without any cytotoxic effect on hMSCs. Interestingly, no adverse effects on hMSCs were detected when the sole temperature change was applied.

Other works exploiting topography changes on PCL and PCL-based substrates have been reported (Table 2), exploring different crosslinking strategies (e.g., thermal crosslinking [69–71], radical crosslinking [72]), microfabrication techniques (e.g., hot embossing [69, 70, 72], replica molding [73], film deformation [71]), and cell types (e.g., 3T3 fibroblasts [69–71], rat bone marrow stem cells (rBMSCs) [72, 73]). As a common trend, from these studies emerges the possibility to actively influence cell and nuclear alignment by exploiting the change in surface topography induced by the change in environmental temperature. In this regard, the temperature explored in the above-mentioned studies ranged between 28 and 38 °C, reported to be a cytocompatible temperature interval. Interestingly, the possibility to guide the differentiation (myogenic [72] (Fig. 6A), adipogenic, or osteogenic [73]) of the cells seeded on the developed micro- and nano-patterned surfaces was also reported.

The SME of PCL-based polymers has been also investigated for other biomedical applications. Gong and co-workers [74] reported the production of UV-crosslinked polyethylene glycol-polycaprolactone (PEG-PCL) microspheres via oil-in-water (o/w) emulsion and their subsequent programming achieved by embedding them into polyvinyl alcohol (PVA) films, deformation with subsequent cooling (60 and 0 °C), and selective dissolution of PVA. Interestingly, the PEG-PCL microspheres displayed reversible, two-way SME when subjected to cyclic heating and cooling between 0 and 43 °C. The microspheres were thus *in vitro* challenged on a mouse microphage cell line, demonstrating different rates of internalization depending on the shape (i.e., higher internalization for spherical vs. ellipsoidal microspheres) and the possibility to achieve intracellular shape-memory recovery, thus opening the floodgates to new strategies towards intracellular drug delivery.

PCL has also been reported for the obtainment of shape-memory foams with a dynamic porous architecture changing during cell cultivation [75]. Such foams were investigated as 3D scaffolds to *in vitro* control the behavior (i.e., cell and nuclear alignment) of MC3T3-E1 preosteoblasts though a temperature change (30 vs. 37 °C).

Another SMP extensively investigated for the fabrication of cell culture surfaces is polyurethane (PU). In this regard, thermoplastic polyurethane (TPU) has been mainly

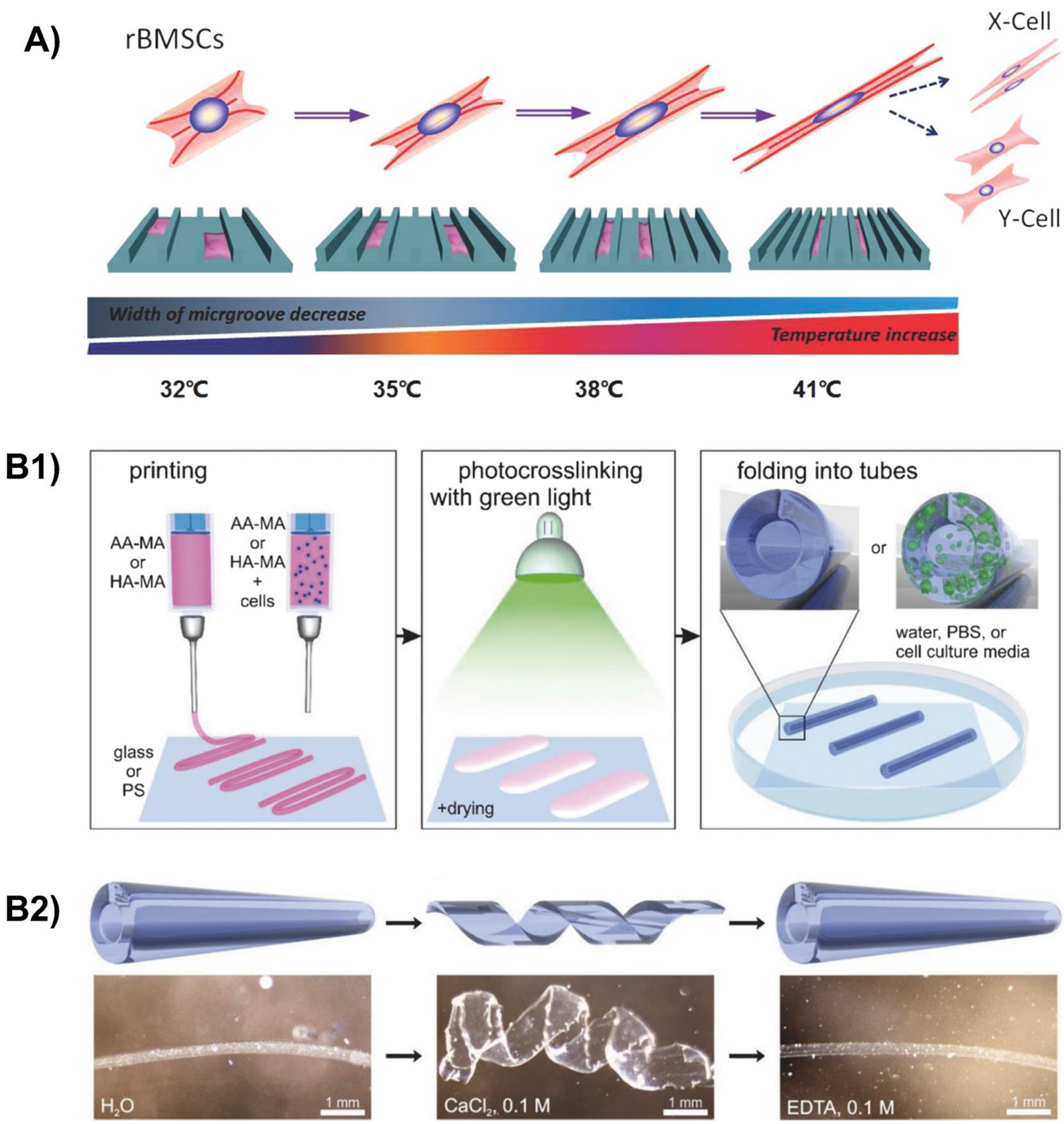

**Fig. 6** 4D fabrication of scaffolds and cell culture surfaces. **A** Programmable control of rBMSCs shape, cytoskeleton reorganization, and differentiation by exploiting the change in microgrooved topography activated by thermally-triggered SME. Reproduced with permission from [72], Copyright (2014), John Wiley and Sons. **B1** 4D fabrication of self-folding AA-MA or HA-MA structures without/with cells (left), crosslinking (530 nm) and mild drying (center), and folding into tubes upon immersion in solution (right). **B2** Reversible shape transformation of AA-MA tubular structure by selective swelling/deswelling in the presence of $Ca^{2+}$ ions (middle) or EDTA (right). Reproduced with permission from [85], Copyright (2017), John Wiley and Sons

reported for the fabrication of scaffolds with controlled fibers architecture via electrospinning, where programming has been achieved through deformation at temperatures higher than the polymer's $T_g$ (60 - 70 °C) [76–78]. Such electrospun scaffolds displayed a $T_g$ close to body temperature, good $R_f$ and $R_r$ values (both > 95 %), and cytocompatibility on different cell lines (human adipose stem cells (hASCs), human fibrosarcoma cell line (HT-1080), and multipotent murine mesenchymal stem cell line (C3H/10T1/2)). Moreover, the change in fibers architecture from aligned to random, induced by temperature increase (30 to 37 °C), led to a change in cell and nuclear alignment without any cytotoxic effect on the seeded cells.

TPU has been also processed via additive manufacturing technology. In this regard, Hendrikson et al. [79] fabricated TPU scaffolds via FFF, programming them via deformation at high temperature (65 °C). hMSCs were then seeded on the scaffolds, that were then cultured at 30 °C to allow cell adhesion and proliferation. The temperature was then set at 37 °C, releasing the strain imparted (during programming) to the scaffolds through the recovery of their permanent shape.

Cell aligned in an elongated shape in response to strain, as well described in the literature [80].

Other SMPs have been investigated for the fabrication of cell culture surfaces. Co-polymers of *tert*-butyl acrylate (*t*BA) and butyl acrylate (BA), UV crosslinked in the presence of a crosslinker and a photoinitiator (triethylene glycol dimethacrylate (TEGDMA) and 2,2-dimethoxy-2-phenyl acetophenone (DMPA), respectively), have been investigated for the production of shape-memory scaffolds to guide the fate of the cells seeded on their surface. The *t*BA-BA co-polymers have been fabricated both in the form of films and foams and deformed (i.e., programmed) before cell seeding. The obtained scaffolds displayed high $R_f$ and $R_r$ values (both > 97 %) and no cytotoxic effects when challenged with hASCs. Similarly to what observed in the studies previously reported in this paragraph, the shape change for these scaffolds was obtained by changing the environmental temperature (from 30 to 37 °C), leading to changes in cell and nuclear alignment [77, 81].

Other SMPs have been also investigated (Table 2), such as poly(propylene glycol) bis(2-aminopropyl ether) and bisphenol A diglycidyl ether, used to fabricate a 4D programmable culture substrate with self-morphing capability (change in surface micropatterns) to induce the differentiation on neural stem cells (NSCs) into neurons and glial cells [82], or PCL-modified polymers, used to fabricate tissue culture surfaces [83]. For further details on the use of SMPs (and, more in general, stimuli-responsive materials) to induce dynamic cell responses, the reader is referred to [84].

**3.4.1.2 SMHs** In the studies described above, SMPs were employed first for cell culture/scaffolds fabrication and cells were only subsequently deposited, according to fabrication approaches 4D-A and 4D-B (Sect. 2.1). 4D bioprinting has recently emerged as a technology combining 3D bioprinting with SMHs, offering the possibility of achieving a shape transformation of the bioprinted construct (i.e., cells + bioink) in response to an applied stimulus. This technology holds great potential for the fabrication of complex and dynamic structures with high resolution, otherwise unattainable with 3D bioprinting technology.

Kirillova and co-workers [85] studied a methacrylated hyaluronic acid hydrogel (HA-MA), photo-crosslinked by green light (530 nm) exposure, for the 4D fabrication of cellularized constructs with shape morphing properties. Specifically, they embedded mouse bone marrow stromal cells (D1 cells) into 3 % HA-MA gels, printed them via an extrusion-based 3D bioprinter, and achieved a shape change of the printed constructs from flat to tubular. Such a shape change was induced by preferential swelling in cell culture medium, due to the presence of crosslinking gradients in the constructs obtained during printing (Fig. 6B1). Interestingly, the same authors also demonstrated the possibility of reversible shape transformation (Fig. 6B2) in another hydrogel, i.e., methacrylated alginate (AA-MA) gel, by controlling the swelling and deswelling process in the presence of $Ca^{2+}$ (additional crosslinker) and EDTA ($Ca^{2+}$ chelating agent). Such reversible shape change can lead to advantages in several biomedical applications, such as loading/release of drugs and cells.

Similarly to this study, Käpilä et al. [86] fabricated swelling-actuated self-folding constructs exploiting crosslinking gradient in shape-changing photo-degradable hydrogels. Poly(ethylene glycol) diacrylate (PEGDA) hydrogels incorporating *ortho*-nitrobenzyl (*o*-NB) moieties were transformed from 2D flat sheets to 3D tubular structures after exposure to UV light (365 nm). Such gels were further functionalized with cell-adhesive peptides (RGD) for both seeding (4D-B) and encapsulation (4D-C) with C2C12 mouse myoblasts. Interestingly, the viability of C2C12 cells was not negatively affected by the UV light-induced folding. Such platforms offer the possibility to provide dynamic, 3D cell culture environments useful to study biological processes sensitive to both physical and temporal cues. Other SMHs have been also investigated for the fabrication of cell culture surfaces and scaffolds. The reader is referred to Table 3 for further details.

### 3.4.2 Bone tissue engineering

**3.4.2.1 SMPs** Bone defects can develop from trauma, infection, congenital etiology, or bone-tumor resection [87]. Large—or critical size—bone defects, i.e., those having a length of the deficiency 2–3 times the bone diameter [88], usually require grafting due to the insufficient bone's self-healing ability. The gold standard technique for critical bone filling is autologous bone graft, even if comorbidity associated with the presence of a second surgical site represent a major disadvantage. Allogenic bone grafts may instead present the risk of disease transmission and are characterized by high costs (e.g., obtainment, treatment, sterilization, and storing) [87]. Bone substitutes, both of biological (e.g., demineralized bone matrix, platelet-rich plasma, bone morphogenic proteins, hydroxyapatite, corals) or synthetic (calcium phosphates, bioactive glasses, polymer-based bone substitutes) origin may represent viable alternatives for large bone defects filling [87].

In this panorama, SMPs, thanks to their shape tunability (useful for minimally invasive surgery), coupled with good mechanical properties, biodegradabilty, and biocompatibility, hold great promise for treating irregular bone defects [89]. Moreover, *in situ* shape recovery of SMPs can also provide a perfect fill of bone defects, ensuring mechanical continuity at the tissue-SMP interface [90].

PCL has been widely investigated in bone tissue engineering field. However, to better control its biodegradation rate and increase its mechanical properties, it is generally modified before processing via i) loading with inorganic fillers, ii) crosslinking, or iii) blending with other polymers [89, 91].

Liu et al. [92] fabricated hydroxyapatite-loaded PCL (PCL-HAp) porous scaffolds, crosslinked via free radical reaction, and loaded with bone morphogenic factors (BMP-2), challenging them both *in vitro* with BMSCs and *in vivo* in mandibular bone defects in a rabbit model. Such scaffolds displayed good shape-memory properties ($R_f$ = 90 % and $R_r$ = 94 %) and a $T_m$ close to body temperature, undergoing shape recovery within few minutes (1 and 10 min *in vitro* and *in vivo*, respectively). The BMP-2-loaded scaffolds displayed no cytotoxic effect *in vitro* and promoted the *in vivo* deposition of new bone in the defect area compared to unloaded (i.e., without BMP-2) scaffolds.

Erndt-Marino and co-workers [93] fabricated a shape-memory foam based on UV photo-crosslinked polycaprolactone diacrylate (PCLDA) as scaffold for irregular bone tissue defects. Interestingly, to avoid compromising the SME adding conventional inorganic fillers (e.g, calcium phosphates and sulfates, HAp, bioactive ceramics), polydopamine (PD) coating was exploited to improve the osteoconductivity of the foams, which *in vitro* promoted the osteoblastic differentiation of hMSCs without enhancing the expression of adipogenic and chondrogenic markers.

Blending with poly(L-lactide) (PLLA) has been reported as an effective strategy to tune the biodegradation rate and the mechanical properties of PCL [94–97]. In this regard, Arabiyat et al. [98] recently reported the fabrication of porous scaffolds made of PCLDA/PLLA semi-interpenetrating network. The scaffolds displayed stiffness values in the range of trabecular bone, accelerated degradation (3 vs 2.5 % weight loss for PCLDA/PLLA and PCLDA, respectively) compared with PCLDA SMPs, and *in vitro* osteoinductive capacity when challenged with hMSCs, demonstrated by the increase in the expression of osteogenic markers (osterix, BMP-4, and collagen 1 alpha 1).

Shape-memory polyurethanes (SMPUs) have been investigated to produce scaffolds for bone tissue engineering. Yang et al. [99] reported biodegradable SMPUs scaffolds capable to promote bone calcification, with significant potential for minimally invasive implantation. They first synthesized a diisocyanate from hexamethylene diisocyanate (HDI) and isosorbide (ISO), then used as a coupling agent in the synthesis of linear SMPUs (ISO-PUs) from poly(DL-lactic acid)-based macrodiol as the soft segment and ISO as the chain extender. The obtained ISO-PUs displayed good SME ($R_f$ = 99.8 % and $R_r$ = 90.2 %), a $T_m$ around body temperature, and high mechanical properties (E = 1000 MPa at 37 °C). Moreover, ISO-PUs completely degraded *in vitro* within 120 days, without cytotoxic effects on rat bone osteoblasts.

Similarly to PCL, inorganic fillers have been used in combination with PU in this field. Xie and colleagues [100] developed a SMP foam based on PU-HAp for the treatment of load-bearing bone defects. Such foams displayed $R_f$ = 94 %, $R_r$ = 91 %, $T_m$ = 40 °C, and self-fitting function (60 s recovery) *in vivo* in a rabbit model. Such properties, coupled with excellent biocompatibility, enhanced bone ingrowth, and promoted neo-vascularization, make PU-HAp foams ideal for minimally invasive bone tissue engineering.

Interestingly, SMPUs have been also explored in combination with Mg particles, promising photothermal fillers, for the production of near-infrared (NIR)-responsive scaffolds. Interestingly, SMPU/Mg composite porous scaffolds were fabricated by low-temperature rapid prototyping (LT-RP) technology, achieving optimal SME ($R_f$ = 93.6 %, $R_r$ = 95.4 %) when irradiated with NIR light (808 nm, 1 W $cm^{-2}$) (Fig. 7A1). Furthermore, SMPU/Mg scaffolds supported *in vitro* cell survival (MC3T3-E1 and murine BMSCs), proliferation, and osteogenic differentiation, while on a rat model they provided tight-contacting and osteopromotive functions (Fig. 7A2) [101].

Other SMPs have been investigated in this field, as biodegradable poly(d,l-lactide-*co*-trimethylene carbonate), fabricated in the form of fibrous scaffolds by electrospinning [102], or poly(butanetetrol fumarate), in the form of porous scaffolds by salt leaching [103]. The reader is referred to Table 2 for better insight.

**3.4.2.2 SMHs** SMHs have also been reported for bone tissue engineering applications, exploiting 4D bioprinting technology.

Lee and co-workers [104] reported the use of oxidized and methacrylated alginate (OMA) and and methacrylated gelatin (GelMA) for the 4D biofabrication of cell-laden bilayered constructs with shape changing capabilities. In particular, they exploited the differences in swelling rations between the two layers to drive structural changes in the printed constructs (Fig. 7B). The developed ink supported normal cellular functions (e.g., adhesion, proliferation) and the differentiation toward osteogenic and chondrogenic lineages with no adverse effects on cell viability.

In another study, Ding and co-workers [105] disclosed the possibility to incorporate an UV absorber directly into photo-crosslinkable inks to create highly tunable crosslinking gradients during the fabrication process. In particular, they investigated three different hydrogels, i.e., OMA, GelMA, and 8-arm PEG-acrylate for the 4D biofabrication via extrusion-based technology of shape changing constructs, incorporating three cell types: hMSC, NIH-3T3 cells, and a cervical cancer cell line (HeLa). Interestingly, OMA-based bioinks allowed for a reversible shape change of the 4D printed constructs simply by controlling the pH and the presence of $Ca^{2+}$ ions/EDTA in solution. All the obtained

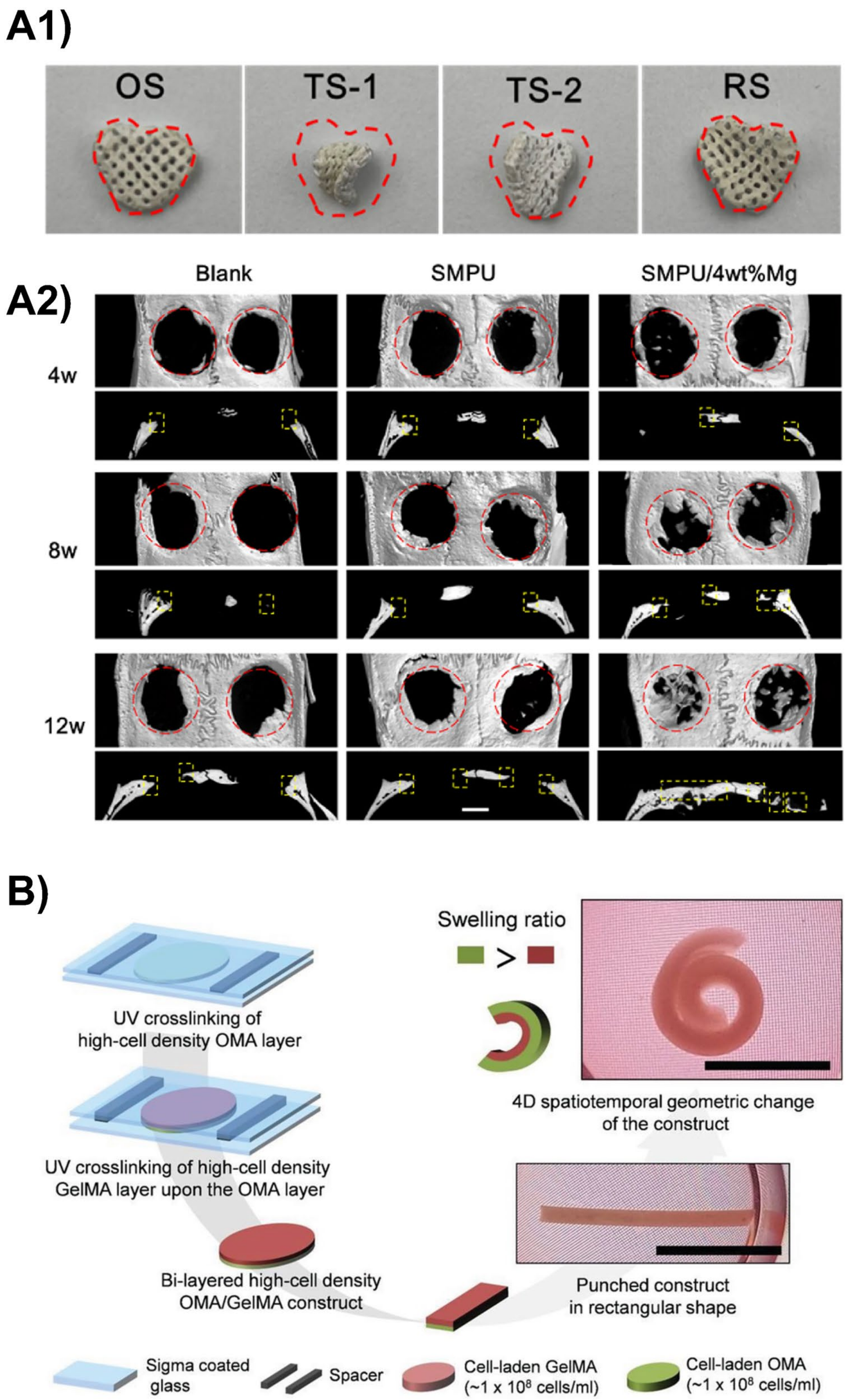

**Fig. 7** 4D fabrication in bone TE. **A1** Shape recovery process of irregular SMPU/Mg composite porous scaffold (4 wt.% Mg) irradiated with 808 nm laser. OS = original shape, TS-1 = Temporary Shape 1; TS-2 = Temporary Shape 2; RS = Recovered Shape. **A2** Micro-CT 3D reconstruction of defective bones and their sagittal images at 4, 8, 12 weeks for *in vivo* animal study. Red circle = defective area. Yellow frame = new bone tissues. Scale bar = 2 mm. Reproduced from [101]. **B** 4D fabrication of cell-laden OMA/GelMA bilayers and shape change from frat to rolled due to the different swelling ratios of the layers. Scale bar = 1 cm. Reproduced with permission from [104], Copyright (2021), John Wiley and Sons

formulations allowed for sustained cell viability (up to 4 weeks). Moreover, an osteogenesis study was conducted on 4D printed hMSC-laden constructs, investigating osteogenic markers such as alkaline phosphates (ALP) activity and calcium deposition, revealing the potentiality of such constructs for bone tissue engineering purposes.

### 3.4.3 Cardiovascular tissue engineering

**3.4.3.1 SMPs** Cardiovascular tissue engineering (CVTE) has emerged as one of the fastest-growing area within the field of tissue engineering, due to the increase of associated diseases (e.g., affecting the cardiac tissue, coronary blood vessels, and valves) and the growing demand for tissue replacement and reconstruction [106, 107].

Traditional CVTE approaches involve the use of 3D scaffolds mimicking the ECM of the tissue, cells, and bioactive molecules (e.g, growth or differentiation factors) to achieve regeneration [108]. More recently, 3D printing and bioprinting have attracted increasing attention in the cardiovascular field, as platforms for the fabrication of complex scaffolds and constructs in layer-by-layer fashion [109]. Despite significant advances in the use of 3D printing and bioprinting in the manufacturing of cardiovascular devices, conformational changes of the printed structure (e.g. taking into account the anatomy of the patient), have not been considered in these approaches. In this regard, 4D printing represents a further step in the fabrication of structures capable to change their shape, function, and properties over time, and can be particularly interesting for cardiovascular devices [110, 111].

**3.4.3.2 Stents and grafts** Polylactic acid (PLA) is a thermoplastic aliphatic polyester with remarkable properties, among which high-strength, biocompatibility, biodegradability, and shape-memory behavior. PLA has been investigated in the field of CVTE to fabricate self-expandable biodegradable vascular stents. In this regard, FFF technology has been investigated for the obtainment of stents, then programmed by compressing them at $T > 60\,°C$ (i.e., $T_g$ of PLA) and fixing the shape at room temperature. Shape recovery was achieved by heating above the $T_g$. Interestingly, such obtained stents displayed optimal shape-memory properties ($R_f$ and $R_r$ values close to 100 %) [112, 113].

Poly(glycerol dodecanoate) acrylate (PGDA) has been investigated in the field of CVTE as it displays a $T_{trans}$ in the 20 - 37 °C range [11]. In particular, shape-memory vascular implants (stent and graft) with mechanical and geometrical adaptability were 4D printed out of PGDA via FFF technology, then crosslinked via coupled UV (365 nm) and thermal (165 °C) treatment. After programming (i.e., compressed temporary shape), the printed implants were *in vivo* implanted into a mouse aorta. The recovery of permanent shape of the implants occurred *in vivo*, after implantation, by exposure to mouse blood flow. Interestingly, such 4D printed vascular structures exhibited high $R_f$ and $R_r$ values (100 and 98 % at 20 and 37 °C, respectively), cycling stability, and rapid recovery time (0.4 s at 37 °C), paving the way to the next generation of vascular implants.

Following a different fabrication approach, Trujillo-Miranda and co-workers [114] proposed highly aligned and self-actuating electrospun bilayers for potential vascular graft applications. In particular, bilayers were obtained in a two-step process, by first depositing a polyhydroxybutyratre (PHB) or PCL layer, followed by a second HA-MA layer. Tubular structures with tunable diameter were obtained by immersing the bilayer structures in aqueous media, which acted as the driving force to achieve the shape transformation. Interestingly, the PHB-bilayer allowed human umbilical vein endothelial cells (HUVECs) culturing without a negative effect on its shape transformation ability. The PHB-based tubular structure demonstrated excellent mechanical stability, superior biocompatibility, and degradability compared to PCL/HA-MA bilayer, thus representing a potential solution for blood vessel replacement.

Overall, 4D fabricated vascular implants hold significant advantages compared to the commercially available ones, which only provide fixed dimensions and mechanical properties.

**3.4.3.3 Cardiac patches and constructs** 4D printing has been reported for the production of cardiac patches with the ability to transform (i.e., tunable architecture) over time, for cardiac tissue regeneration. On this topic, Miao and co-workers [115] fabricated 4D thin films (< 300 $\mu$m) with hierarchical micropatterns using a photolithographic-stereolitographic-tandem strategy (PSTS) starting from natural lipids (i.e., soybean oil epoxidized acrylate, SOEA). As expected, hMSCs cultured on the fabricated structures attached to the grooves, spread, and expanded, lastly aligning in the groove direction. Moreover, hMSCs cultured on the micropatterned PSTS films underwent cardiomyogenic differentiation. Interestingly, the obtained films self-bended after exposure to a thermal stimulus (i.e., 37 °C), as a consequence of the crosslinking density gradient formed during the photolithographic process. Overall, due to their potential easy integration with damaged tissues or organs, the obtained structures lend themselves well as 4D patches for cardiac tissue regeneration.

In another approach, smart cardiac constructs acting both as minimally invasive cell vehicles and *in situ* tissue patches have been proposed for the regeneration of damaged myocardial tissue [116, 117]. In particular, Wang et al. [117] fabricated NIR-responsive 4D cardiac constructs in a two-step process. First, they fabricated micro-patterned molds via DLP technology from photo-crosslinkable PEGDA. Then, the molds were filled with an ink made of bisphenol A diglycidyl ether (BDE), poly(propylene glycol) bis(2-aminopropyl ether) (PBE), decylamine (DA), and graphene nanoplatelets (GNPs) to fabricate the 4D constructs. Human induced pluripotent stem cell-derived cardiomyocytes (hiPSC-CMs), hMSCs, and human endothelial cells (hECs) were co-cultured on the 4D constructs presenting aligned microgrooves and adjustable curvature, leading to a uniform

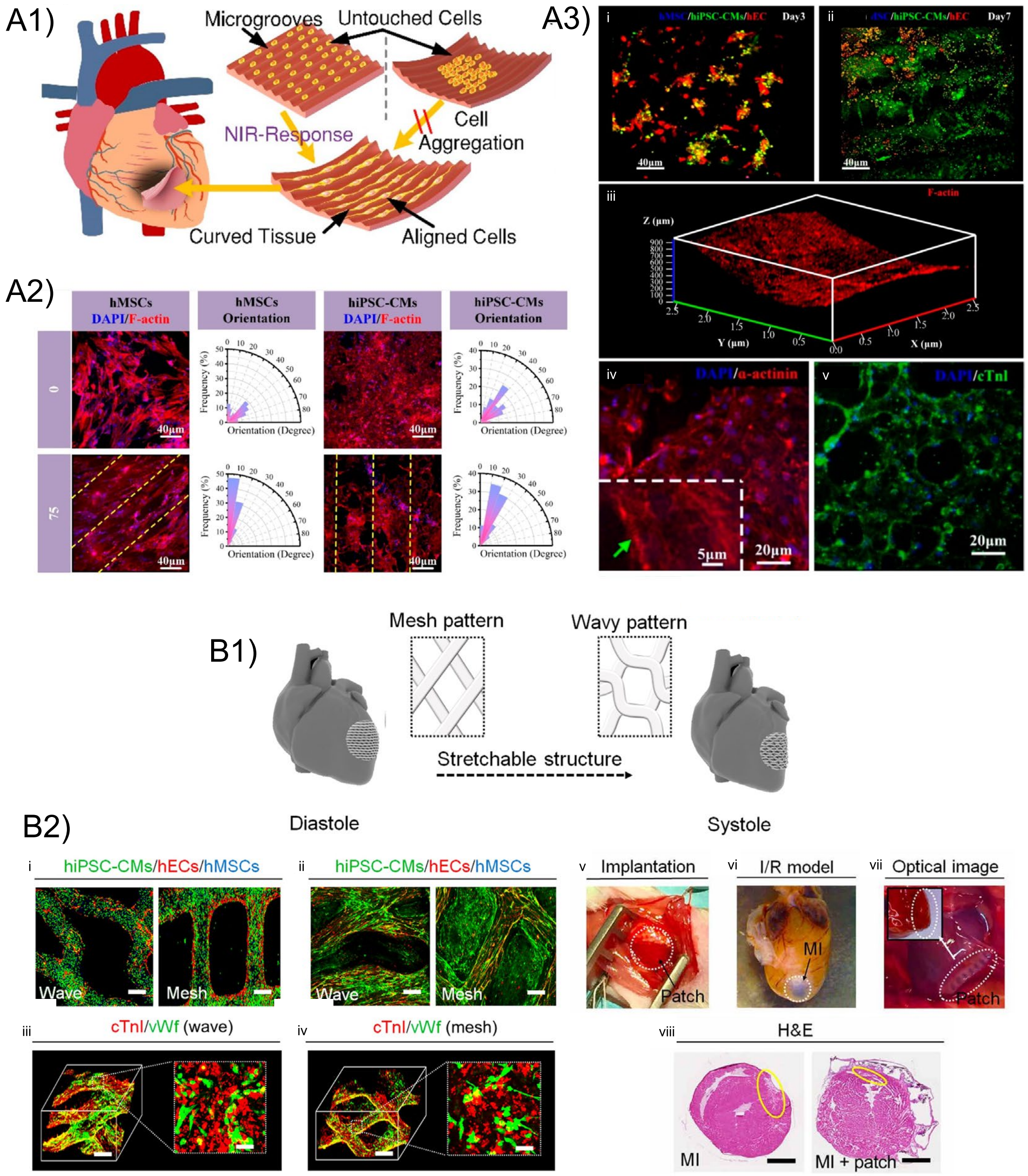

**Fig. 8** 4D fabrication in cardiovascular TE. **A1** cell-laden 4D printed cardiac patch for myocardial infarction treatment and its shape change from flat to curved. **A2** Morphology and orientation of hMSCs and hiPSC-CMs on microgrooves with different widths (0 vs. 75 $\mu$m) 7 days after cell seeding. Dashed yellow lines = orientation of the microgrooves. **A3** Immunofluorescence images of the 4D cardiac constructs: co-culture of hMSCs, hECs, and hiPSC-CMs at (i) day 3 and (ii) day 7; (iii) F-actin staining displaying aligned and uniformly distributed cells on the curved surface of 4D constructs; nuclei and (iv) $\alpha$-actinin or (v) anti-cardiac troponi (cTnI) of cardiac cells cultured on the 4D patches. Reproduced with permission from [117], Copyright (2021), American Chemical Society. **B1** CAD models of the 3D architecture during the cardiac cycle. **B2** Immunofluorescence images of the cellularized 4D printed patches: tri-cultured hiPSC-CMs, hECs, and hMSCs at (i) 1 day and (ii) 7 days of culture (scale bars = 200 $\mu$m); cTnI and vascular protein (vWf) on the (iii) wave-patterned and (iv) mesh-patterned patches (scale bars = 200 µm for 3D images and 20 $\mu$m fro 2D insets); optical images of (v) surgical implantation of the patch, (vi) heart MI model, and (vii) implanted patch at week 3; H&E images of mouse hearts without treatment (MI) or patch-treated (MI + patch) at week 10 (yellow circles = infarct area; scale bars = 800 µm). Reproduced from [118]

distribution of aligned cells and excellent myocardial maturation on the curved constructs (Fig. 8A). More recently, the same formulation (BDE, PBE, DA, GNPs) was investigated for the fabrication of a 4D thermo-sensitive cardiac construct for myocardial regeneration [116]. Temperature ($T_{trans}$ = 43 °C) was here exploited to induce a shape change in the printed structure, from minimally invasive cell vehicles to *in situ* tissue patches. Interestingly, the fabricated 4D cardiac constructs displayed an optimal shape-memory behavior ($R_r$ = 100 %) and, when challenged with hiPSC-CMs, an outstanding myocardial maturation.

**3.4.3.4 SMHs** Additionally to the SMPs described above, SMHs have also been studied for CVTE. In this regard, Cui and co-workers [118] fabricated a hydrogel-based 4D cardiac patch with physiological adaptability (Fig. 8B). GelMA/PEGDA hydrogel was used to fabricate the 4D structures via SLA. The shape transformation from flat to bend was then achieved exploiting the crosslinking gradient present in the 4D structures, coupled with swelling. Interestingly, it was found that the shape change process allowed to achieve 3D conformations nearly identical to the physiological surface curvature of the heart. The obtained structures were then tri-cultured with hiPSC-CMs, hMSCs, and hECs to obtain a complex cardiac tissue. The *in vivo* maturation of the 4D printed cellularized patches was also evaluated into a murine model of chronic myocardial infarction. Interestingly, three weeks after implantation the 4D printed patches exhibited excellent engraftment and vascularization. In addition, histological analysis revealed a decrease in the infarcted area treated with the 4D patches compared to the untreated control, indicating the high regenerative potential of the patches. Overall, this study revealed the possibility to reproduce the anisotropy of elastic epicardial fibers and vascular networks, as well as guiding contracting cells for engineered cardiac tissue.

In a different investigation, Pedron et al. [119] presented a strategy for cardiac microtissue transplantation using hydrogel/polymer bilayers capable to roll or unroll at will. In details, the bilayer was fabricated through photolithographic processes from diacrylated triblock copolymer layer composed of poly(ethylene glycol)/poly(lactic acid) (PLA-b-PEG-b-PLA) and poly(methyl methacrylate) (PMMA), coupled with poly(N-isopropyl acrylamide) (PNIPAM) hydrogel layer, well-known for its thermo-responsive character. In aqueous environment, PNIPAM shows significant changes in its swelling rate at temperatures close to its lower critical solution temperature (LCST), causing the bilayer structures to roll or unroll in response to minor changes in temperature. H9C2, A431 rat cardiac cell lines, and primary neonatal rat cardiomyocytes (RCm) were then cultured on such bilayers when in the flat conformation. At confluency, the rolling-up of the bilayer constructs was achieved by slight temperature decrease, leading to cell delivery constructs preserving intercellular interactions. Such constructs hold the potential to unroll in response to a temperature increase (i.e., 37 °C) in the selected injured site (e.g., myocardium), thus offering the possibility of a smart delivery of cardiac microtissues.

Exploiting the same mechanism of the previous study, Liu and co-workers [120] reported the fabrication, via DIW, of acellularized (i.e., apporach 4D-A) gel tubes composed of an active thermo-responsive gel, PNIPAM and a passive (i.e., non-responsive) gel, polyacrylamide (PAAm). Assisted by finite element modeling, the authors fabricated tubes with different periodic vertical and horizontal arrangements of active and passive segments, generating a wide range of shape changes including uniaxial elongation, bending, and radial expansion. Interestingly, given the transition temperature close to body temperature (i.e., 34 °C), such tubular structures lend themselves well for applications in the fields of vascular tissue engineering.

### 3.4.4 Neural tissue engineering

**3.4.4.1 SMPs** The nervous system is a complex, highly structured network of cells responsible for regulating the functions and activities in our body. When injury occurs, neural tissue experiences changes in its complex architecture. In favorable circumstances, damaged cells (i.e., axons) can regrow re-establishing connections with their targets, as in the case of non-critical (< 1-2 cm [121]) injuries in peripheral nerves. In contrast, central nervous system axons typically fail to regenerate [122]. In this scenario, traumatic brain injury, neurodegenerative diseases, spinal cord injury, and critical peripheral nerve injury, involving the disruption of axonal pathways or tracts with consequent loss of structure and functions of the neural tissue, are some of the major causes of neuro-disability. Fortunately, the treatment of neural tissue injuries has entered a new era thanks to advances in tissue engineering and regenerative medicine techniques [122, 123].

**3.4.4.2 Neural scaffolds and nerve guidance conduits** Focusing on 4D fabrication, Miao and co-workers fabricated micro-patterned nerve guidance conduits through SLA-based printing, using SOEA ink loaded or not with GNPs [124]. Interestingly, the thus fabricated conduits displayed one-way shape-memory behavior undergoing programmed deformation from flat to folded upon thermal trigger ($T_g$ = 20 °C). Interestingly, the conduits also displayed two-way shape-memory behavior, reversibly changing their shape (flat to folded) after exposure to different solvents (i.e., water and ethanol). In addition, hMSCs cultured on the 4D printed conduits underwent neural differentiation. Overall, given their noteworthy characteristics including physical guidance and possibility of self-entubulation, the proposed conduits could dynamically and seamlessy integrate into the stumps of a damaged peripheral nerve. In another paper from the same research group, a cell culture substrate capable to recapitulate the complex topographic changes associated with the neurodevelopment process over time [82] was fabricated combining FFF, SLA, and thermomechanical imprint technologies. The 4D culture substrate, made from BDE, PBE, and DA, exhibited a thermal-triggered ($T_g$ = 37 °C) self-morphing process over time, responsible for the regulation of NSCs behavior, i.e., alignment and neural differentiation. Overall, the proposed substrate was capable of replicating the physiological characteristics of NSC-derived neural development, also offering the possibility of deepening current knowledge on neural tissue regeneration or to mimicking the progression of specific neurological diseases.

**3.4.4.3 Brain tissue modeling** 4D bioprinting has emerged as a noteworthy approach for fabricating dynamic, responsive *in vitro* brain tissue models capable to recapitulate the architecture and folding biomechanics of brain tissue, to study the effects that these parameters have on the neurodevelopment process. In particular, a great interest has been devoted to exploring how the cortical tissue of the brain enfolds to generate its convoluted surface, as already discussed in previous review papers [36, 125]. In this regard, an illustrative study where 4D printing has been exploited for the obtainment of a brain model was reported by Cui et al. [126]. They synthesized a NIR-responsive nanocomposite (BDE + PBE + DA + 16 % GNPs), exploiting photothermal-triggered shape-memory behavior to dynamically and remotely control the spatio-temporal transformation of the printed structure (Fig. 9). Interestingly, the printed structured offered electroconductive and optoelectronic properties, which allowed the NSCs seeded on their surface to undergo growth and neurogenic differentiation.

**3.4.4.4 SMHs** Despite the significant improvements of 4D bioprinting technology, the development of 4D biofabricated constructs for neural tissue engineering is still in its infancy. In this regard, to the best of these authors' knowledge, no works exploiting SMHs for the 4D biofabrication of neural constructs can be found in the literature. A possible explanation relies on the fact that current 4D bioprinting techniques face highly complex challenges. This is even more true in neural TE, where 4D bioprinting success relies closely on the development of optimal formulations (i.e., inks) based on SMHs for the fabrication of constructs mimicking native tissue [125, 127].

### 3.4.5 Muscle tissue engineering

**3.4.5.1 SMPs** Muscle tissue constitutes approximately 45 % of the mass of an adult human body and is responsible for all dynamic activities, from movement (e.g., locomotion, eye movement) to metabolism regulation. Traumatic injuries, pathological events, and surgery (e.g., tumor removal) are among the prevalent reasons for reconstructive muscle surgery. In this panorama, muscle TE has emerged as a strategy to generate engineered tissues capable of restoring muscular normal functions or replacing defective muscles [128]. Current technologies, like 3D (bio)printing fail to recapitulate the dynamic mechanical cues (e.g., stretching, folding) capable to guide muscle cell fate and lead to myogenic alignment and functional maturation. 4D fabrication can take up this challenge, generating structures (scaffolds and constructs) capable of undergoing programmed changes in shape and properties over time [36, 60].

Miao and co-workers [129] combined FFF and surface coating techniques to understand how topographical cues can guide the commitment of hMSCs towards skeletal muscle type. In particular, they 4D fabricated scaffolds made from PCL/SOEA with excellent strain fixity and recovery rates ($R_f$ and $R_r$ = 96 and 100 %, respectively). The obtained scaffolds underwent rapid shape change by a thermal trigger (T = 37 °C), and the topographical cues led to enhanced expression of myogenic proteins and genes (e.g., myoblast differentiation protein-1, desmin, and myosin heavy chain-2), suggesting their suitability for muscle tissue regeneration.

PCL was also used in combination with AA-MA for the 4D fabrication of a bilayer scaffold using electrospinning. The obtained bilayers underwent shape transformation (triggered by $Ca^{2+}$ ions) from flat to scroll-like tubular structure upon exposure to an aqueous buffer. Interestingly, the self-folding process allowed to encapsulate myoblasts (C2C12 cells), previously seeded on the flat scaffold. The myoblasts were further shown to align in the direction of the PCL fibers and differentiate into aligned myotubes capable to contract when electrically stimulated (Fig. 10A) [130].

**3.4.5.2 SMHs** As in the case of the last work presented above, SMPs have been reported in combination with hydrogels to achieve an overall SME. Nevertheless, SMHs have also been studied alone for muscle TE purposes. In this regard, Vannozzi et al [131] designed bilayers coupling PEGDA hydrogels with two different molecular weights. The bilayer, obtained via photolithography, underwent self-folding due to the differential swelling ratios of each layer. C2C12 and Cor4U-human iPS cell-derived cardiomyocytes were seeded on the unfolded flat bilayer, then self-rolling occurred by incubation (37 °C, cell culture medium), leading to cell encapsulation. Cell viability was confirmed 72 h after incubation, indicating that the self-folded PEGDA scaffold allowed for optimal nutrients and oxygen exchange to all cells. Overall, these platforms are promising systems to be used as implantable tissue building blocks.

More recently, Yang and co-workers [132] proposed an electrically-assisted 3D printing approach for the 4D biofabrication (i.e., approach 4D-C) of a skeleton muscle model using a GelMA-based bioink. Specifically, C2C12 cells-laden microfibers were fabricated combining 3D printing technology and an electric field, in an approach recognized in the literature as cell electrowriting (CEW) [133]. Using an optimized set of process parameters (e.g.,

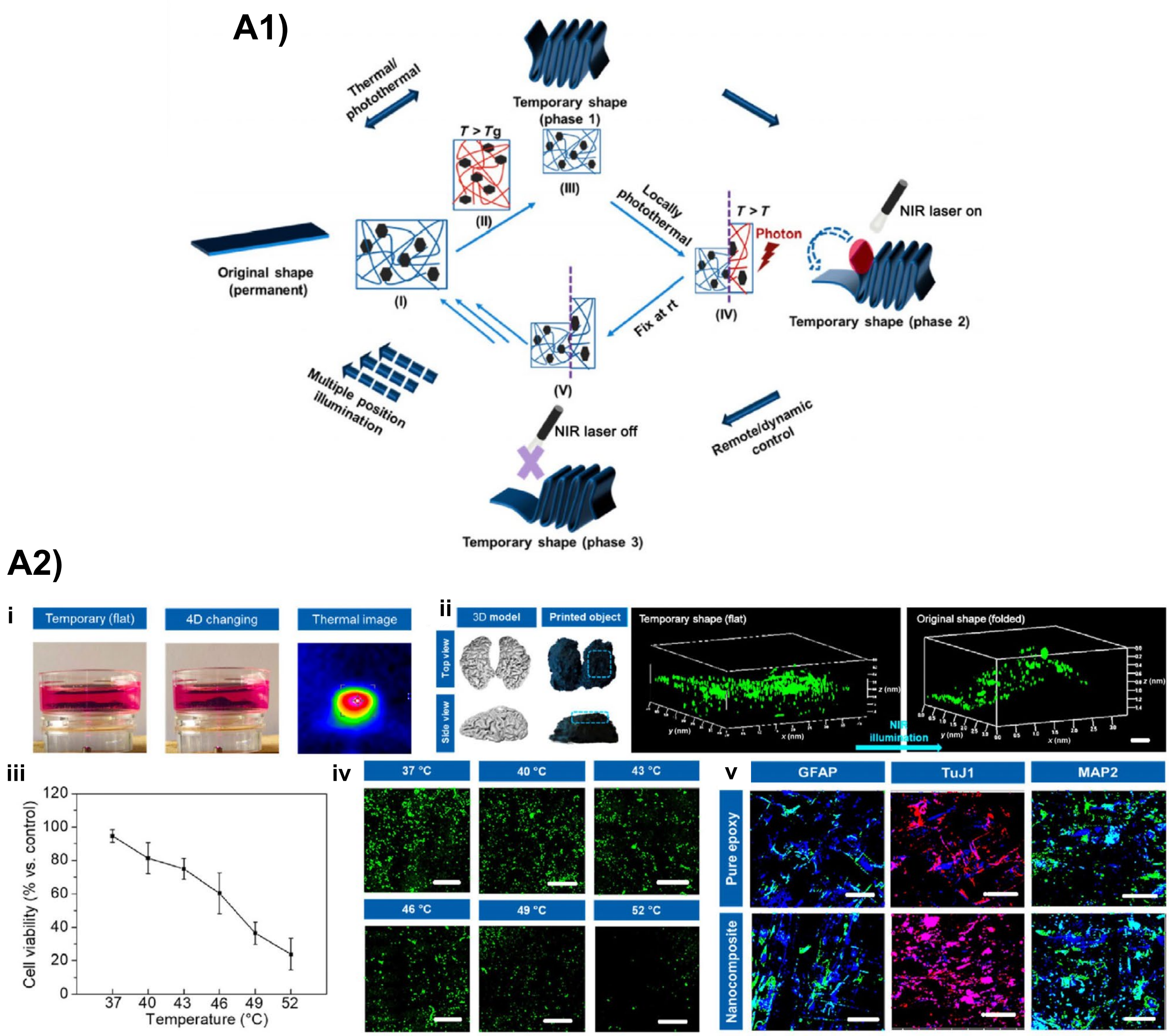

**Fig. 9** 4D fabrication in neural TE. **A1** NIR-induced transformation of 4D-printed nanocomposite: (I) permanent shape, (II) temporary shape obtained after NIR irradiation ($T > T_g$) applying an external force, (III) fixing at room temperature, (IV and V) gradual heating to $T_g$ under NIR exposure to control the shape-changing position and transformation time (phases 2 and 3). **A2** NIR-responsive 4D printed brain construct: (i) 4D transformation and thermal image of brain constructs under NIR irradiation; (ii) GFP-NSCs distribution on 4D brain constructs during shape change (scale bar = 500 $\mu$m); (iii) NSC viability as a function of temperature; (iv) fluorescent images of GFP-NSCs at different temperatures (scale bar = 200 $\mu$m); (e) immunofluorescence images of NSC differentiation on 4D printed brain construct compared to control (pure epoxy construct) after 2 weeks of cell culture (scale bar = 200 μm). Reproduced with permission from [126], Copyright (2019), Springer Nature

electric field density, time, cell number), the electric field allowed to induce cell alignment and myogenic differentiation. Moreover, a shape change from flat to tubular was obtained on thus obtained microfibers through exposure to cell culture medium, suggesting their potential use as muscle models for *in vitro* testing purposes (Fig. 10B).

### 3.4.6 Tracheal tissue engineering

**3.4.6.1 SMPs** The trachea is a cartilaginous conduit which connects the larynx to the bronchi, providing warm, humid, and clean air to the lungs, clearing secretions and keeping the airway free. Tracheal replacement is necessary after critical resection (i.e., resection a tracheal segment longer than 6 cm in adults) or in case of injury. In this regard, a conventional solution is allogeneic transplantation, which is, however, associated with the shortcomings of immunosuppressant therapy and the severe lack of healthy donors. Thus, TE has come to the limelight as a potential strategy to tackle this clinical problem, proposing tubular scaffolds with the potential to remodel and vascularize without the risk of rejection [134]. With the progress of additive manufacturing technology, it is nowadays possible to design personalized tracheal models suited to the patient's anatomical specifications [135, 136]. However, scaffold loosening and fracture represent frequent causes of failure. With the aim of tackling such constrains, 4D printing has emerged for the fabrication of adaptive structures capable to ideally fit the trachea and provide optimal fixation.

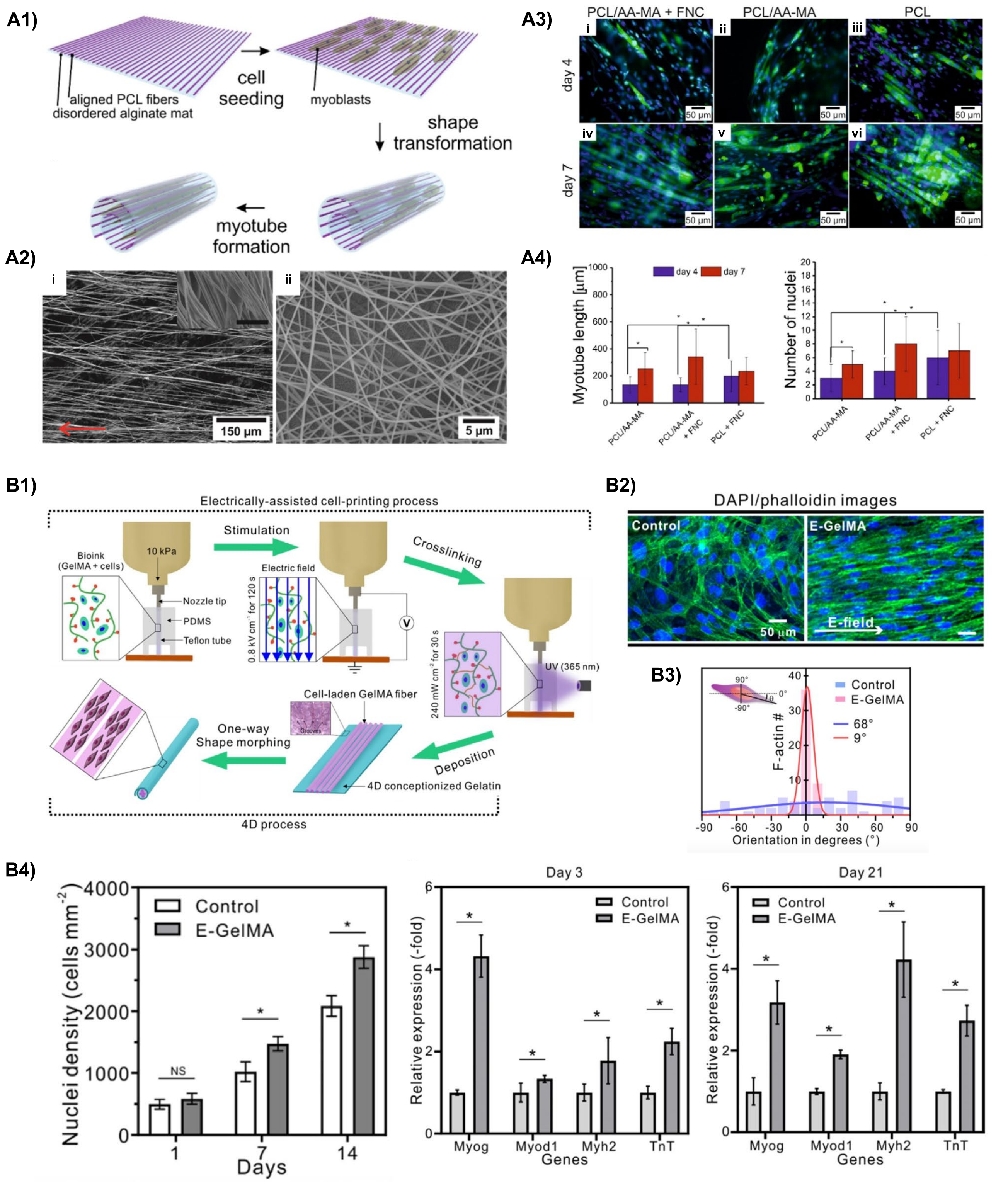

**Fig. 10** 4D fabrication in muscle TE. **A1** PCL/AA-MA bilayer mats fabricated via electrospinning, C2C12 cells seeding, and shape transformation. **A2** SEM images of (i) aligned PCL and (ii) random AA-MA fibers (red arrow indicates the fiber direction). **A3** Immunofluorescence images (green = myosin heavy chain, blue = nuclei) showing myogenesis in C2C12 muscle cells at day 4 and 7 of differentiation on the bilayers (FNC = fibronectin coating) **A4** length and number of nuclei after 4 and 7 days of differentiation. Reproduced from [130]. **B1** 4D biofabrication via cell electrowriting (CEW) of shape morphing muscle fiber-like structures. **B2** Fluorescence images of cells in control and GelMA (white arrows = electric field direction). **B3** Orientation angle of F-actin. **B4** Cell nuclei density after 1, 7, and 14 days of cell culture and expression of myogenin (Myog), myogenic differentiation 1 (MyoD1), myosin heavy chain 2 (Myh2), and Troponin T (TnT) genes after 3 and 21 days of culture. Reproduced from [132]

In this regard, Pandey et al. [137] explored DIW to 4D fabricate tracheal scaffolds starting from thermo-responsive PLA/PCL blend. Depending on the composition (i.e., PLA weight ratios: 30 - 100 %), the obtained scaffolds displayed a transition temperature in the 61.5 - 49.1 °C range. The PLA70/PCL30 blend was selected as optimal, due to its remarkable shape-memory properties ($R_f$ and $R_r$ = 91 and 90 %, respectively). When tested *ex-vivo* in a goat trachea model, the programmed scaffold was able to regain its original shape (folded to flat), optimally fitting to the lumen of the trachea (Fig. 11A).

Another approach proposed in the literature consists in FFF-based fabrication of tracheal stents/scaffolds using $Fe_3O_4$ nanoparticles-loaded PLA, possessing magnetothermal shape-memory behavior [138, 139]. In particular, increased $Fe_3O_4$ nanoparticles amounts (5 - 18 wt. %) resulted in scaffolds with increased mechanical properties and increased magnetothermal effect. The addition of the nanoparticles did not affect the $T_g$ (~ 65 °C) of the fabricated structures, which underwent fast shape recovery ($R_r$ > 99 %) upon heating ($T \sim T_g$) induced by alternating magnetic field in the 30–50 kHz range.

**3.4.6.2 SMHs** In addition to the SMPs described above, SMHs have also been studied for tracheal TE purposes. In this regard, Kim and co-workers [140] proposed a methacrylated silk-fibroin (Sil-MA) hydrogel for the 4D biofabrication of tracheal substitutes. They biofabricated Sil-MA hydrogel bilayers via DLP bioprinting, combining condrocites (for tracheal cartilage side) and turbinate-derived mesenchymal stem cells, TBSCs (for respiratory mucosa side). The 4D shape transformation from flat to hollow tubes was then achieved simply by immersion in cell culture medium, exploiting the different swelling ratios between the two layers (flat vs. patterned). The thus obtained 4D constructs were *in vitro* cultured for 3 days, then implanted into a rabbit tracheal damage model. Results showed that the constructs were fully integrated within the host tissue, and both epithelium and cartilage were formed at the defect sites (Fig. 11B). This work suggested the potentiality of 4D biofabrication in the reconstruction of a damaged tissue, which offers the possibility to obtain constructs with shape-morphing ability in mild conditions, great reliability, and biocompatibility. Such a work can be considered pioneering in the field and among the most advanced related to 4D biofabrication (one of the few at the pre-clinical, *in vivo* research stage). Further work in this direction could certainly open the floodgates to possible future clinical applications of 4D bioprinting.

## 3.5 Conclusions

### 3.5.1 Potentialities and limitations

In this work, an overview of 4D fabrication for TE purposes was proposed, with a focus on shape-changing soft materials. In particular, SMPs and SMHs investigated in the literature for the 4D fabrication of cellularized structures have been thoroughly reviewed.

Undoubtedly, 4D fabrication is gaining increasing importance in the field of TE, playing a key role in the fabrication of structures capable of shape transformation in response to environmental stimuli, thus mimicking the dynamic behaviors of native tissues and offering unprecedented control over tissue regeneration processes. This technology represents a shift in perspective compared to traditional TE approaches, in which the structures (e.g., scaffolds, constructs) and devices are instead generally "passive", therefore unable to adapt to changes in biological environments. In this work, it has been depicted how the obtainment of 4D constructs facilitates the development of biomimetic tissue models with enhanced functionality and physiological relevance. By harnessing cells and shape-memory materials, researchers can create complex, multi-cellular architectures that closely resemble native tissue structures. In addition, the tunability of the 4D fabricated constructs allows for precise control over their mechanical, chemical, and biological properties, enabling the design of advanced and customized TE solutions. This level of control is essential for achieving desired outcomes in the diverse fields of application seen in this work.

Despite significant progress, several challenges remain to be addressed to fully realize the potential of 4D fabrication in TE. Hereafter, some limitations and possible strategies are described.

**3.5.1.1 4D fabrication approaches** The main disadvantage in the use of SMPs for 4D fabrication purposes is that cells are generally seeded on their surface after the fabrication process: this mean that only 4D-A and 4D-B approaches are accessible to obtain cellularized constructs. Indeed, harsh environmental conditions for cells, like high temperatures

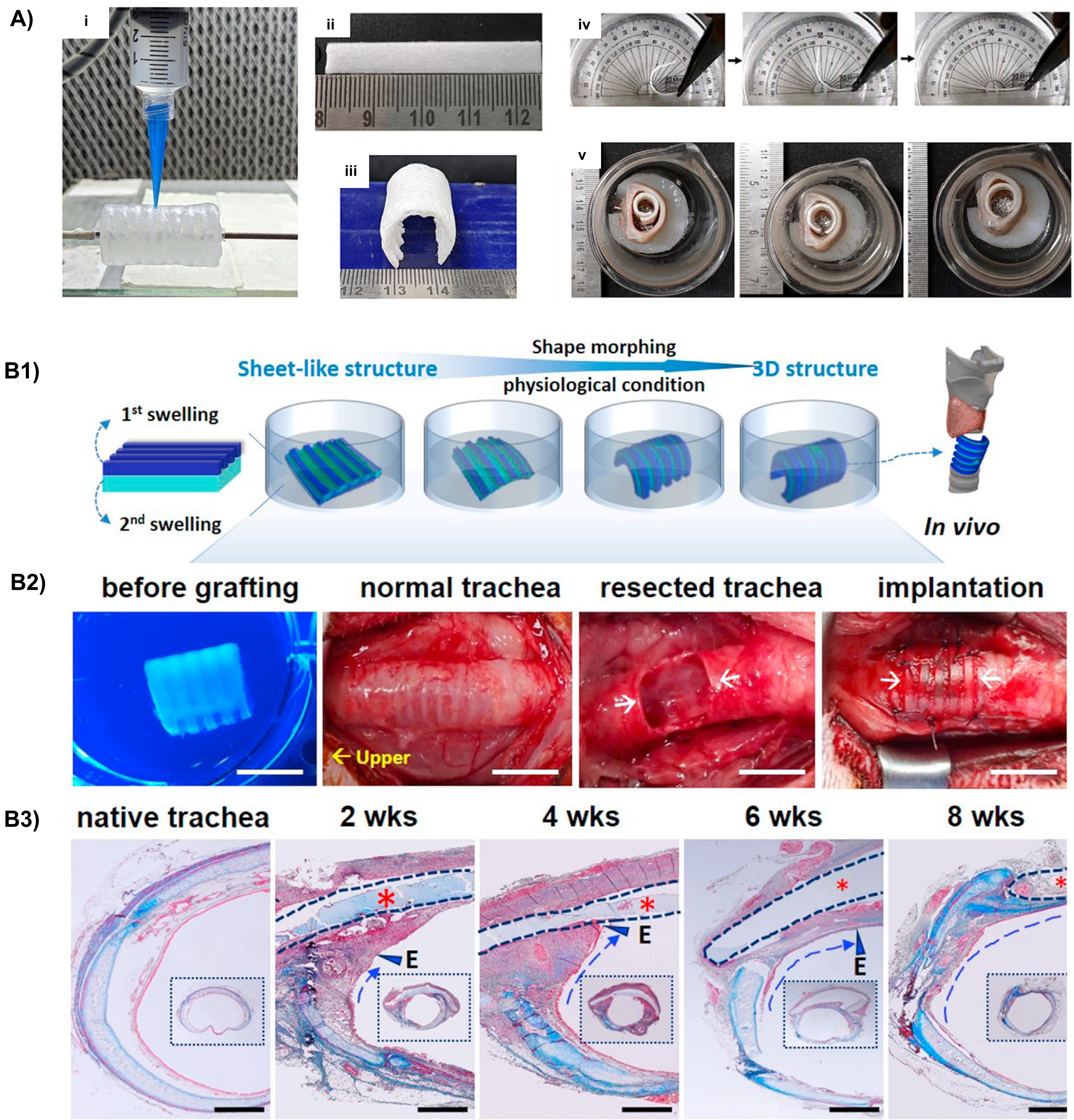

**Fig. 11** 4D fabrication in tracheal TE. **A** 4D printing of PCL/PLA using DIW technology: (i) 3D printed scaffold, (ii) flat scaffold, and (iii) tubular scaffold; (iv) thermally activated SME of tracheal scaffold (iv) in air or (v) in hot water. Reproduced with permission from [137], Copyright (2022), Elsevier. **B1** Shape change of 4D printed Sil-MA hydrogels induced by swelling due to osmotic pressure. **B2** Transplantation of the 4D bioprinted trachea into a damaged rabbit trachea (scale bars = 1 cm). **B3** Masson's trichrome staining of native trachea and 4D bioprinted trachea 2, 4, 6, and 8 weeks after transplantation. Red asterisks and dotted line = 4D bioprinted trachea (scale bars = 1 mm). E and dotted line with arrow head = region of regenerated epithelium. Reproduced with permission from [140], Copyright (2020), Elsevier

(>> 37 °C), cytotoxic solvents, or high electric fields used during processing prevent exploiting the 4D-C approach with SMPs. On this topic, SMHs take up the challenge of being employed for all three approaches, i.e., 4D-A, B, C. This is made possible by the cell-friendly conditions employed during the fabrication and post-processing steps

**Table 2** 4D fabricated SMPs for tissue engineering (TE)

| Material(s) | Bio-degradability | Stimulus | $R_f$ (%), $R_r$ (%) | Approach | Fabrication technique | Shape change | Cell type | Application | Ref(s) |
|---|---|---|---|---|---|---|---|---|---|
| NOA-63 | – | T ($T_g \sim 37\,°C$) | – | 4D-B | Hot embossing | Topography | C3H/10T1/2 | Cell culture | [156] |
| PCL | – | T ($T_m = 36.2\,°C$) | 99, 98 | 4D-B | Hot embossing | Topography | hMSCs | Cell culture | [68] |
| PCL | – | T ($T_m = 33\,°C$) | 99, 90 | 4D-B | Hot embossing | Topography | 3T3 | Cell culture | [69] |
| TPU | Yes | T ($T_g = 48\text{-}49\,°C$) | 99–100, 94–96 | 4D-B | Electrospinning | Fibers architecture | hASCs | Cell culture | [76] |
| tBA-BA | – | T ($T_g = 40\,°C$) | >97, > 97 | 4D-B | Injection molding | Surface wrinkling | hASCs | Cell culture | [81] |
| PCL/AuNRs | – | Photothermal ($T_m = 38\,°C$) | 99, 94 | 4D-B | Hot embossing | Topography | 3T3 | Cell culture | [157] |
| PCL/AL | Yes | T ($T_m = 38{,}41\,°C$) | 94.2, 96.8 (41 °C) | 4D-B | Hot embossing | Topography | rBMSCs | Cell culture | [72] |
| PCL | – | T ($T_m = 33\,°C$) | 99, 90 | 4D-B | Hot embossing | Topography | 3T3 | Cell culture | [70] |
| 6A PEG-PCL | Yes | T ($T_m = 38{,}41\,°C$) | 96.4, 92.4 (41 °C) | 4D-B | Replica molding | Topography | rBMSCs | Cell culture | [73] |
| tBA-BA | – | T ($T_g = 42\,°C$) | – | 4D-B | Salt leaching | Pores architecture | hASCs | Cell culture | [77] |
| TPU | Yes | T | [76] | 4D-B | Electrospinning | Fibers architecture | hASCs | Cell culture | [77] |
| TPU | – | T ($T_g = 32.2\,°C$) | – | 4D-B | FFF | 3D architecture | hMSCs | Cell culture | [79] |
| PCL | – | T ($T_m = 33\,°C$) | 98.1, 89.9 | 4D-B | Injection molding | Topography | 3T3 | Cell culture | [71] |
| TPU | – | T ($T_g = 48\,°C$) | 99, 99 | 4D-B | Electrospinning | Fibers architecture | HT-1080, C3H/10T1/2 | Cell culture | [78] |
| SMPU-imHA | – | T ($T_m = 37\,°C$) | – | 4D-B | Water foaming | Pores architecture | MC3T3-E1 | Cell culture | [75] |
| PCL | Yes | T ($T_m = 54\,°C$) | 100, 93 | 4D-B | Injection molding | 3D architecture | L929 | Cell culture | [83] |
| Gelatin/PCL, Gelatin/PHF | Yes | T/solvent | – | 4D-B | Dip-coating | 3D architecture | Primary fetal mouse neural stem cells | Cell culture | [158] |
| *6A PEG-PCL | Yes | T ($T_{high} = 43\,°C$) | > 90, > 90 | 4D-B | o/w emulsion | Microspheres shape | Mouse macrophage cell line | Phagocytosis | [74] |
| SOEA | – | T ($T_g = 20\,°C$) | 92–99, ~ 100 | 4D-A | Stereolithography | 3D architecture | hMSCs | Scaffold | [159] |
| PCL/castor oil | – | T ($T_g = -8\text{-}35\,°C$) | 92–100, ~ 100 | 4D-A | Casting | 3D architecture | hMSCs | Scaffold | [160] |
| SADs | Yes | T ($T_g = 18.3–26.8\,°C$) | –, > 90 | 4D-A | DLP | 3D architecture | NOR-10 | Scaffold | [161] |
| *PCL/PNIPAM | – | Solvent | – | 4D-A | Electrospinning | 3D architecture | 3T3 | Scaffold | [162] |
| PU/PEO/ Gel/ SPIO NPs | Yes | T ($T_m = 48.8–62.1\,°C$) | 100, 100 | 4D-A | LFDM | 3D architecture | hMSCs | Bone TE | [163] |
| TPU | – | T/solvent (37 °C) | 67, 90 | 4D-A | SLS | 3D architecture | MG-63 | Bone TE | [164] |
| SMPU/Mg | – | Photothermal ($T_m = 58.1\,°C$) | 93.6, 95.4 | 4D-A | LT-RP | 3D architecture | MC3T3, rBMSCs | Bone TE | [101] |
| PLGA-g-PCL/ PPDLDA | – | T ($T_m = 37\,°C$) | 97, 87 | 4D-A | Salt leaching | 3D architecture | BMSCs | Bone TE | [165] |
| PFP/PCL | Yes | T ($T_m = 37\,°C$) | 97, 98 | 4D-A | FFF | 3D architecture | MC3T3-E1 | Bone TE | [166] |
| PDLLA-co-TMC | Yes | T ($T_g = 36.7, 44.2\,°C$) | > 94, > 98 | 4D-A | Electrospinning | 3D architecture | Primary rat osteoblasts | Bone TE | [102] |
| PBF | Yes | T/solvent ($T_g = 130\,°C$) | > 95, > 95 | 4D-A | Salt leaching | 3D architecture | Primary rat osteoblasts | Bone TE | [103] |
| PCL | – | T | – | 4D-A | Salt leaching | 3D architecture | hBMSCs | Bone TE | [93] |
| PCL/HAp | Yes | T ($T_m = 37\,°C$) | 90, 94 | 4D-A | Microparticles leaching | 3D architecture | rBMSCs | Bone TE | [92] |
| PCL/PLLA | Yes | T | – | 4D-A | Salt leaching | 3D architecture | hMSCs | Bone TE | [98] |
| PU | Yes | T ($T_g = 42\,°C$) | > 99.8, > 90.2 | 4D-A | Injection molding | 2D architecture | Rat osteoblasts | Bone TE | [99] |
| PU/HAp | Yes | T ($T_g = 40\,°C$) | > 94, > 91 | 4D-A | Gas foaming | 3D architecture | MC3T3-E1 | Bone TE | [100] |
| PU/HAp | Yes | T ($T_g = 40–50\,°C$) | 90, 96 | 4D-A | Replica molding | 3D architecture | rMSCs | Bone TE | [167] |

**Table 2** (continued)

| Material(s) | Bio-degradability | Stimulus | $R_f$ (%), $R_r$ (%) | Approach | Fabrication technique | Shape change | Cell type | Application | Ref(s) |
|---|---|---|---|---|---|---|---|---|---|
| TPU/HAp | Yes | T ($T_g = 43\,°C$) | 92.8, 93.4 | 4D-A | FFF | 3D architecture | Primary fibroblasts | Cartilage TE | [168] |
| SOEA | – | T | – | 4D-A | Photo/stereolithography | 2D architecture | hMSCs | Cardiac TE | [115] |
| PEGDA | – | T ($T_{trans} = 20\text{-}37\,°C$) | – | 4D-A | FFF | 3D architecture | – | Cardiac TE | [11] |
| PEGDA | – | T ($T_{trans} = 20\text{-}37\,°C$) | – | – | FFF | 3D architecture | – | Cardiac TE | [11] |
| PLA | Yes | T ($T_g = 90\,°C$) | > 99, > 99 | – | FFF | 3D architecture | – | Cardiac TE | [112] |
| PLA | Yes | T ($T_g = 66\,°C$) | –, > 97 | – | FFF | 3D architecture | – | Cardiac TE | [113] |
| CA-PLA-PEG | Yes | T ($T_g = 37\,°C$) | 96, > 94 | 4D-A | FFF | 3D architecture | L929 | Cardiac TE | [169] |
| BADGE | – | Photothermal ($T_g = 45\,°C$) | – | 4D-B | Replica molding | 3D architecture | hiPSC-CMs, hMSCs | Cardiac TE | [117] |
| BADGE | – | T ($T_g = 36\,°C$) | –, ~100 | 4D-A | Replica molding | 3D architecture | hiPSC-CMs, hMSCs | Cardiac TE | [116] |
| PCL/HA-MA, PHB/HA-MA | Yes | Solvent | – | 4D-B | Electrospinning | 3D architecture | HUVECs | Vascular TE | [114] |
| PCL/SOEA | – | T ($T_{trans} = 37\,°C$) | 96, ~100 | 4D-A | Casting | 3D architecture | hMSCs | Muscle TE | [129] |
| PCL/AA-MA | Yes | Solvent/ions | – | 4D-B | Electrospinning | 3D architecture | C2C12 | Muscle TE | [130] |
| PCL-PU/HA-MA | Yes | Solvent | – | 4D-B | DIW/MEW | 3D architecture | C2C12 | Muscle TE | [170] |
| **SOEA (graphene) | – | Solvent/T | – | 4D-A | SLA | 3D architecture | hMSCs | Neural TE | [124] |
| BADGE | – | T ($T_g = 37\,°C$) | – | 4D-B | Hot embossing | Topography | NSCs | Neural TE | [82] |
| PCL-PGS/HA-MA | Yes | Solvent | – | 4D-B | Electrospinning | 3D architecture | PC-12 | Neural TE | [171] |
| BADGE/GNPs | – | Photothermal ($T_g = 45\,°C$) | 99, 95 | 4D-B | FFF + extrusion | 3D architecture | NSCs | Neural TE | [126] |
| PLA/PCL | – | T ($T_{trans} = 49.1\text{–}61.5\,°C$) | 91–98, 74–90 | – | DIW | 3D architecture | – | Tracheal TE | [137] |
| $PLA/Fe_3O_4$ | – | Magnetothermal ($T_g = 65\,°C$) | – | – | FFF | 3D architecture | – | Tracheal TE | [139] |
| $PLA/Fe_3O_4$ | – | Magnetothermal ($T_g = 66\,°C$) | –, > 99 | – | FFF | 3D architecture | – | Tracheal TE | [138] |

4D-A = fabrication of a non-cellularized scaffold, shape change, and seeding with cells; 4D-B = fabrication of a non-cellularized scaffold, cell seeding, and shape transformation of the construct; 4D-C = biofabrication of a cellularized construct and shape change

** = two-way SME

Table abbreviations: 3T3 = mouse embryonic fibroblasts; *AA-MA* methacrylated alginate, *AL* allyl alcohol, *AuNRs* gold nanorods, *BADGE* bisphenol A diglycidyl ether, *C2C12* mouse myoblast cell line, *C3H/10T1/2* mouse embryonic fibroblasts, *CA-PLA-PEG* cinnamic acid - polylactic acid - polyethylene glycol co-polymer, *Gel* gelatin, *GelMA* methacrylated gelatin, *GNPs* graphene nanoparticles, *HA-MA* methacrylated hyaluronic acid, *hASCs* human adipose stem cells, *hMSCs* human mesenchymal stem cells, *HAp* hydroxyapatite, *hiPSC-CMs* human-induced pluripotent stem cell-derived cardiomyocytes, *HUVECs* human umbilical vein endothelial cells, *HT-1080* human fibrosarcoma cell line, *imHA* isocyanate-modified hydroxyapatite, *L929* murine fibroblast cell line, *LFDM* low-temperature fuse deposition manufacturing, *LT-RP* low temperature rapid prototyping, *MC3T3-E1* mouse preosteoblast cell line, *MEW* melt electrowriting, *MG-63* human osteosarcoma cell line, *NOR-10* mouse skeletal muscle fibroblasts, *NSCs* neural stem cells, *PEO* polyethylene oxide, *PBF* poly(butanetetrol fumarate), *PC-12* pheochromocytoma neuronal cell line, *PCL* polycaprolactone, *PCS* poly(glycerol sebacate), *PDLLA-co-TMC* poly(L-lactide-co-trimethylene carbonate), *PEGDA* poly(glycerol dodecanoate) acrylate, *PFP* poly(propylene fumarate), *PHF* copolymer of hecanediol and fumaryl chloride, *PLGA-g-PCL/PPDLDA* poly(L-glutamic acid)-g-poly (ϵ-caprolactone) co-polymerized with acryloyl chloride grafted poly(ω-pentadecalactone), *PLLA* poly(L-lactide), *PNIPAM* poly(N-isopropylacrylamide), *PU* polyurethane, *rBMSCs* rabbit bone marrow stem cells, *SADs* salicylic acid derivatives, *SOEA* soybean oil epoxidized acrylate, *SPIO NPs* iron oxide nanoparticles, *tBA-BA* tert-butyl acrylate and butyl acrylate co-polymer, *rBMSCs* rat bone marrow stem cells, *RT* room temperature, *TPU* thermoplastic polyurethane

(e.g., physiological temperature, aqueous environment, non-toxic crosslinking). Interestingly, 4D-C approach simplifies the obtainment of cellularized contructs, representing a one-pot process for their obtainment, significantly reducing processing times and steps.

**Table 3** 4D fabricated SMHs for tissue engineering (TE)

| Material(s) | Bio-degradability | Stimulus | Actuation mechanism | Approach | Fabrication technique | Shape change | Cell type | Application | Reference(s) |
|---|---|---|---|---|---|---|---|---|---|
| **PAN | Yes | Solvent/ions | Ion-triggered crosslinking | 4D-B | Casting | 3D architecture | hMSCs | Cell culture | [172] |
| o-NB PEGDA | – | Light | o-NB photodegradation | 4D-B, 4D-C | Casting | 3D architecture | C2C12 | Cell culture | [86] |
| *AA-MA, HA-MA | Yes | Solvent/ions | Swelling (crosslinking gradient) | 4D-B, 4D-C | DIW | 3D architecture | D1 | Scaffold/construct | [85] |
| *PEGDA, PEGDA/GelMA | Yes | Solvent | Swelling (different MWs, material) | 4D-A, 4D-C | Photolithography | 3D architecture | MDA-MB-231, SUM159-GFP | Scaffold/construct | [173] |
| PEGDA | – | Solvent | Swelling (different MWs) | 4D-C | Photolithography | 3D architecture | L929, $\beta$-TC-6 | Scaffold/construct | [66] |
| GelMA, PEGA8, **OMA | Yes | Solvent/ions | Swelling (crosslinking gradient), ion-triggered crosslinking | 4D-C | Photolithography, DIW | 3D architecture | hMSCs, 3T3, HeLa | Bone TE | [105] |
| GelMA, OMA | Yes | Solvent | Swelling (different material) | 4D-C | Gel casting, DIW | 3D architecture | 3T3, hASCs | Bone and cartilage TE | [104] |
| *PNIPAM/(PLA-b-PEG-b-PLA) | Yes | T ($T_{trans}$ = 26 °C) | Sol–gel transition | 4D-B | Photolithography | 3D architecture | H9C2, A431, RCm | Cardiac TE | [119] |
| GelMA, PEGDA | Yes | Solvent | Swelling (crosslinking gradient) | 4D-A | SLA | 3D architecture | hiPSC-CMs, hMSCs, hECs | Cardiac TE | [118] |
| PEGDA | – | Solvent | Swelling (different MWs, thickness) | 4D-B | Photolithography | 3D architecture | hiPSC-CMs, C2C12 | Muscle TE | [131] |
| GelMA | Yes | Solvent | Swelling (different grooves, thickness, crosslinking) | 4D-C | CEW | 3D architecture | C2C12 | Muscle TE | [132] |
| *PNIPAM/PAAM/Laponite | – | T ($T_{trans}$ = 34 °C) | Sol–gel transition | 4D-A | DIW | 3D architecture | – | Vascular TE | [120] |
| *Sil-MA | Yes | Solvent | Swelling (different geometries) | 4D-C | DLP | 3D architecture | Human chondrocytes, TBSCs | Tracheal TE | [140] |

4D-A = fabrication of a non-cellularized scaffold, shape change, and seeding with cells; 4D-B = fabrication of a non-cellularized scaffold, cell seeding, and shape transformation of the construct; 4D-C = biofabrication of a cellularized construct and shape change

** = two-way SME

Table abbreviations: *β-TC-6* mouse insulinoma cells, *A431* human epidermoid carcinoma cell line, *AA-MA* methacrylated alginate: C2C12 = mouse myoblast cell line, *CEW* cell electrowriting, *D1* mouse bone marrow stromal cells, *GelMA* methacrylated gelatin, *H9C2* rat BDIX heart myoblast cell line, *HA-MA* methacrylated hyaluronic acid, *hASCs* human adipose stem cells, *hECs* human endothelial cells, *HeLa* human cervical cancer cell line, *hiPSC-CMs* human-induced pluripotent stem cell-derived cardiomyocytes, *hMSCs* human mesenchymal stem cells, *MDA-MB-231* triple negative breast cancer cell line, *MWs* molecular weights, *o-NB* ortho-nitrobenzyl moieties, *OMA* oxidized and methacrylated alginate, *PAAM* polyacrylamide, *PEGA8* 8-arm PEG-acrylate, *PEGDA* poly(glycerol dodecanoate) acrylate, *PLA-b-PEG-b-PLA* diacrylated triblock copolymer of poly(ethylene glycol) and poly(lactic acid), *PNIPAM* poly(N-isopropylacrylamide), *RCm* primary neonatal rat cardiomyocites, *Sil-MA* methacrylated silk fibroin, *SUM159-GFP* mesenchymal triple-negative breast cancer cell line labeled with green fluorescent protein

**3.5.1.2 Programming the shape-memory behavior** Another limitation is the need for manual or machine-assisted programming of shape-memory materials, that could limit their application when dealing with biological samples. As an example, thermally-triggered SMPs commonly used in 4D printing usually need a thermo-mechanical programming step after fabrication to exhibit shape-memory behavior. A possible strategy to overcome this limitation has already been introduced in this work (Sect. 3.3) and consists in exploiting internal stresses generated in the material during the fabrication process. On this topic, some recent works have unveiled the possibility of programming SMPs during 4D printing, in an approach referred to as direct 4D printing, consisting in generating and "trapping" pre-strains in the printed structures, subsequently recovered by heating. For further details, the reader is referred to [141–143]

**3.5.1.3 Reversibility of shape transformation** Another limitation is the shortage of reversible (i.e., two-way) SME in the majority of the 4D fabricated systems, which need to be reprogrammed each time (into a temporary shape) before shape recovery [36]. Actually, this is not considered a limitation when one-way shape-memory behavior is needed (e.g., for scaffolds/implants adapting to defects sites), but may represent a huge limitation when a reversible SME is needed (e.g., for soft actuators [54] or *in vitro* dynamic models mimicking biological environmental changes [82]).

The leading strategy to tackle this limitation consists in the design of new materials with reversible SME suitable for 4D fabrication. Many efforts are currently being made in this direction by researchers all over the world. The reader is referred to [54, 144, 145] for further details.

### 3.5.2 Future perspectives

Future developments of 4D fabrication in the TE field are envisaged to be driven by three major needs: i) new materials, ii) emerging fields of applications, iii) preclinical/clinical translation.

From a material perspective, the need for new stimuli-responsive materials is critical for the progress in the field. In this regard, particular attention must be paid on the development of biocompatible and biodegradable stimuli-responsive materials [40, 146]. As an example in this field, shape-memory composites (SMCs) and multi-functional SMHs are coming to the limelight as promising materials in 4D fabrication. SMCs combine the advantages of composites (i.e., polymers + filler(s)) with shape-memory behavior, contributing to the improved performance of 4D shape changes or opening new possible applications. In this regard, it has been demonstrated in this work how graphene doping can play an important role when incorporated in 4D inks, acting as a photo-absorbent and generating internal stress gradients inside the printed structures [124]. Additionally, it has been demonstrated to provide the doped inks with NIR-responsive character [117]. Again, the presence of inorganic species, like $Fe_3O_2$ nanoparticles, has been reported to confer the resulting composite with additional magneto-responsive character [147]. On the other hand, multi-functional SMHs, integrating shape-memory behavior with additional functionalities, such as self-healing capabilities [148–150], are anticipated to drive future advancements in the field. In fact, they hold the potential to be tailored for the repair and regeneration of functional tissues, streamlining the implantation process for various tissue geometries and orchestrating the spatio-temporal distribution of diverse cell types.

Furthermore, the development of multi-responsive materials, i.e. capable to respond to multiple stimuli, even physiological or pathological ones (e.g., enzymes [151], glucose levels [152], inflammation [153]), mimicking the body's regulatory mechanisms, could enable the development of increasingly more patient-specific therapeutic approaches.

Another need that could drive future advancement in the field is the targeting of different (i.e., from those described in this work) tissues and organs or even tackling new therapeutic strategies. As an example of the latter point, recent advances of 4D printing in cancer research have been recently reported elsewhere [154]. In this regard, 4D printing represents a cutting-edge strategy for enhancing treatments on targeted sites thanks to the dynamic nature of the fabricated systems and the possibility to engineer and tune their responses, bringing the therapeutic process to the next level.

Finally, significant efforts must be devoted to the preclinical investigation of 4D fabricated structures, as a test bed for their effective clinical translation. In this regard, the understanding of the impact of 4D fabricated structures on the host tissues (e.g., interaction with the immune system) and of the host tissue on the implanted structures (e.g., *in vivo* functionality) is still limited. Ongoing research will help answer this open questions, towards the future translation of 4D technology into clinical practice, which nowadays remains a distant prospect [155]

In conclusion, 4D fabrication holds immense promise for revolutionizing the field of TE. With continued research and innovation, it is poised to drive thrilling advancements in the years to come.

Has this revolution already begun?

**Funding** Open access funding provided by Università degli Studi di Pavia within the CRUI-CARE Agreement. The funding was provided by HORIZON EUROPE European Research Council (Grant No. 101039467). Views and opinions expressed are however those of the author(s) only and do not necessarily reflect those of the European Union or the European Research Council. Neither the European Union nor the granting authority can be held responsible for them. are no conflicts to declare.

## Declarations

**Conflict of interest** There are no conflicts to declare.